\documentclass[11pt]{article}
\pdfoutput=1
\usepackage{soul}
\usepackage{jheppub}
\usepackage{amsfonts}
\usepackage{amsmath}
\allowdisplaybreaks[4]         
\usepackage{amssymb}   
\usepackage{euscript}       
\usepackage{color}          
\usepackage{tensor}     
\usepackage{caption}
\usepackage{subcaption}

\newcommand{\req}[1]{(\ref{#1})} 

\newcommand{\bea}{\begin{eqnarray}}
\newcommand{\eea}{\end{eqnarray}}
\newcommand{\ba}{\begin{eqnarray}}
\usepackage{braket}
\newcommand{\ea}{\end{eqnarray}}

\newcommand{\beq}{\begin{equation}}
\newcommand{\eeq}{\end{equation} }
\newcommand{\beqa}{\begin{eqnarray}}

\newcommand{\eeqa}{\end{eqnarray}}
\newcommand{\beqar}{\begin{eqnarray*}}
\newcommand{\eeqar}{\end{eqnarray*}}

\newcommand{\be}{\begin{equation}}
\newcommand{\ee}{\end{equation}}

\renewcommand{\req}[1]{(\ref{#1})}

\title{ \boldmath Black Hole Multipoles  in Higher-Derivative Gravity}

\author[a]{Pablo A. Cano,}
\author[b]{Bogdan Ganchev,}
\author[a]{Daniel R. Mayerson}
\author[c, d]{and Alejandro Ruip\'erez}

\affiliation[a]{Instituut voor Theoretische Fysica, KU Leuven.
Celestijnenlaan 200D, B-3001 Leuven, Belgium \vspace{0.1cm}}

\affiliation[b]{Institut de Physique Th\'eorique, Universit\'e Paris Saclay, CEA, CNRS, Orme des Merisiers, Gif sur Yvette, 91191 CEDEX, France \vspace{0.1cm}}

\affiliation[c]{Dipartimento di Fisica e Astronomia, Universit\`a di Padova, via Marzolo 8, 35131 Padova, Italy \vspace{0.1cm}}

\affiliation[d]{INFN, Sezione di Padova, via Marzolo 8, 35131 Padova, Italy}

\date{\today}
\abstract{
We consider a broad family of higher-derivative extensions of four-dimensional Einstein gravity and study the multipole moments of rotating black holes therein. 
We carefully show that the various definitions of multipoles carry over from general relativity, and compute these multipoles for higher-derivative Kerr using the ACMC expansion formalism.
We obtain the mass $M_{n}$ and current  $S_{n}$ multipoles as a series expansions in the dimensionless spin; in some cases we are able to resum these series into closed-form expressions.
Moreover, we observe the existence of intriguing relations between the corrections to the parity-odd multipoles $S_{2n}\neq 0$ and $M_{2n+1}\neq 0$ that break equatorial symmetry, and the parity-preserving corrections that only modify $S_{2n+1}$ and $M_{2n}$. Further, we comment on the higher-derivative corrections to multipole ratios for Kerr, and we discuss the phenomenological implications of the corrections to the multipole moments for current and future gravitational wave experiments.
 }

\begin{document} 
\maketitle
\flushbottom

\section{Introduction}
\label{sec:Introduction}

The advent of gravitational wave observations at LIGO and VIRGO \cite{Abbott_2009} have begun to test general relativity in new energy regimes to unprecedented levels of precision. The future space-based observatory LISA \cite{Danzmann_1996} and third-generation ground-based detectors such as the  Einstein Telescope  and Cosmic Explorer \cite{Kalogera:2021bya} will further extend our observations both in energy scales and precision. 

An obvious question to ask in the context of these gravitational wave observations is: \emph{What do we expect to see beyond general relativity (GR)?} One possible avenue to attack this question is to wonder how gravitational wave merger events would be altered if black holes were ``replaced'' by a new type of horizonless exotic compact object (ECO) \cite{Cardoso:2019rvt}. Would we be able to see effects of the lack of horizon? How compact can such objects be, before they are indistinguishable from black holes? Such questions have been examined in gravitational phenomenology from many different avenues, see e.g. \cite{Cardoso:2019rvt,Mayerson:2020tpn,Maggio:2021ans}.

Another, more systematic method of examining what is possible beyond GR is an effective field theory (EFT) approach, which we follow here. It amounts to introducing higher-derivative operators in the gravity sector of the Lagrangian that are suppressed by powers of a new small length scale (or large energy scale). These can include four-derivative gravitational terms such as in dynamical Chern-Simons (dCS) theory \cite{Jackiw:2003pm,Yunes:2009hc,Sopuerta:2009iy} or Einstein-dilaton-Gauss-Bonnet (EdGB) \cite{Pani:2009wy,Pani:2011gy,Kleihaus:2011tg} --- for these,  additional scalars are needed to couple to the four-derivative terms if they are to be dynamically relevant. At higher orders, we can also include six- and eight-derivative operators (without scalars); these have been constructed and considered in e.g. \cite{Endlich:2017tqa,Cardoso:2018ptl}. In such an EFT approach, the relevant question becomes: How well are we able to (or will be able to) constrain the new length scale associated to these irrelevant EFT operators using precision gravitational wave observations? In this paper, we consider the most general possible four-, six-, and eight-derivative corrections in four-dimensional gravity theory.

When higher-derivative corrections are present, the evolution of a binary black hole system will be slightly altered compared to GR --- both due to its corrected coupling to the emitted gravitational wave radiation, and due to finite-size effects --- such as alterations of the multipole structure of the individual black holes. These multipoles are determined by the metric of an individual, isolated Kerr black hole and how this metric is deformed in the presence of higher-derivative corrections. 
The full perturbation of the Kerr solution due to any four-, six-, or eight-derivative corrections was calculated in \cite{Cano:2019ore, Cano:2020cao}.\footnote{Thermodynamical aspects of the corrected solutions were also studied in \cite{Reall:2019sah}.} In this paper, we provide a comprehensive and exhaustive analysis of the multipole structure of this higher-derivative-deformed Kerr solution. Such a comprehensive analysis is new, although certain partial results exist \cite{Cardoso:2018ptl}. Our analysis also corrects the erroneous claim that dCS theory introduces a non-zero $S_4$ multipole for (deformed) Kerr \cite{Sopuerta:2009iy}.

Defining and calculating multipoles in a higher-derivative theory of gravity is subtle. First of all, there are three ways multipoles can be defined for stationary vacuum solutions of four-dimensional GR: the Geroch-Hansen formalism \cite{Geroch:1970cd,Hansen:1974zz}; Thorne's ACMC formalism \cite{Thorne:1980ru}; 
and using the covariant phase-space formalism, as recently done in \cite{Compere:2017wrj}. These three definitions of multipoles are equivalent for stationary vacuum solutions to GR \cite{1983GReGr..15..737G,Compere:2017wrj}; however, it is a priori not clear these methods all generalize to higher-derivative gravity solutions, nor that they would all give equivalent definitions of the multipole moments. We discuss and check that all methods are well defined and give the same results for the gravitational multipoles of the higher-derivative Kerr black hole. Additionally, we show that perturbative field redefinitions do not affect these multipoles, so that this ambiguity present in higher-derivative gravity theories does not affect the physical and observable multipoles.

The higher-derivative corrections to Kerr are known as a complicated perturbative expansion in the rotation parameter $\chi=a/M$. In addition to calculating the multipoles as a perturbative expansion in $\chi$, for certain multipoles we are able to resum this expansion into a closed-form functional expression of $\chi$. These resummed functions give additional insight into the full functional form of the metric in terms of $\chi$.

The Kerr solution has a unique set of multipole moments, completely determined by its mass and angular momentum. By contrast, in string theory, black holes can exhibit a much richer multipole structure. The Kerr solution can then be obtained as a limit of such string theory black holes \cite{Chow:2014cca,Bena:2020uup}. Curiously, one can then consider \emph{ratios} of multipoles in this limit; some of these ratios could be ill-defined for Kerr itself (as its odd-parity multipoles $M_{2\,n+1}$ and $S_{2\,n}$ vanish), but still have a well-defined Kerr limit of string theory black hole multipole ratios. These multipole ratios have been argued to provide constraints on corrections to GR coming from string theory \cite{Bena:2020see,Bena:2020uup}. We examine these multipole ratios within our framework of higher-derivative corrections and comment on the case where these can be related to coming from string theory compactified on a torus, as well as the more general case.

\medskip

In the following Section \ref{sec:summary}, we give a summary of our main results.
Section \ref{sec:HDG} gives a quick overview of the higher-derivative theories we are considering, including arguments to show that these are really the most general ones possible (under certain assumptions and up to eight derivatives), together with a review of the higher-derivative-deformed Kerr black hole solution.
Then, Section \ref{sec:multipoleexp} discusses carefully the three definitions of multipoles and their equivalence in our higher-derivative theories. We put everything together in Section \ref{sec:BHmultipoles}, where we calculate the perturbation to all multipoles for Kerr in the presence of the various higher-derivative deformations; we also briefly discuss the multipole ratios introduced in \cite{Bena:2020see,Bena:2020uup}. Finally, we discuss the observability of the higher-derivative length scale --- through the multipole-related finite-size effects --- in Section \ref{sec:obs}.
Appendix \ref{app:redefinitions} contains further details on the invariance of multipoles under higher-derivative metric field redefinitions, and Appendix \ref{app:Waldformalism} discusses aspects of the Wald formalism for surface charges.

\subsection{Summary}\label{sec:summary}
This paper provides a comprehensive and complete overview of the multipole structure of the Kerr black hole deformed by the most general possible four-, six- and eight-derivative corrections to four-dimensional general relativity. We show that the three existing methods of calculating multipoles in a gravity theory --- the Geroch-Hansen formalism, Thorne's ACMC formalism, and the covariant phase-space formalism --- are all non-trivially well-defined and give equivalent results for higher-derivative-deformed Kerr; we additionally show that higher-derivative field redefinitions do not affect the gravitational multipoles.

We give the explicit expressions for these multipoles --- in Section \ref{sec:mult6D} (see especially eq. (\ref{eq:zn6Derv})), Section \ref{sec:mult8D} (see especially eq. (\ref{eq:Znquartic})), and Section \ref{sec:mult4D} (see especially eq. (\ref{Znquadratic})); additional expressions are given in the ancillary \texttt{Mathematica} notebook on the arXiv version of this paper. We have calculated these multipoles as a perturbative expansion in the rotation parameter $\chi=a/M$ to a high order and expect this expansion to be valid even for fairly large $\chi\sim 1$. Where we were able, we have also resummed these expansions into closed-form functions of $\chi$ --- see (\ref{eq:6Df2full})-(\ref{eq:6Df4full}) and (\ref{hnexact}). The six- and eight-derivative corrections to the multipoles reveal a surprising relation between odd-parity corrections and even-parity corrections --- captured in the complex expressions (\ref{eq:zn6Derv}) and (\ref{eq:Znquartic}).
Further, curiously, the numerical coefficients of these multipoles grow with $n$, indicating a breakdown of the EFT at a certain maximal $n_\text{max}$. We have also investigated how the multipole ratios of \cite{Bena:2020see,Bena:2020uup} are altered by higher-derivative corrections, and how our results are consistent with the conjecture given therein that string theory compactified on a torus predicts a unique value for these multipole ratios.

Finally, we have discussed the observability of the higher-derivative corrections through the finite-size effects of their modifications to the mass multipole $M_2$. The best-case order of magnitude constraint on the higher-derivative length scale $\ell$, which will be accessible in the future with observations at third-generation ground-based detectors such as the Einstein Telescope, is estimated to be:
\be \ell \lesssim 0.1-1\, \text{km} ,\ee
which is roughly two orders of magnitudes better than existing constraints (based on current gravitational wave detections) on this length scale.
We discuss and derive this constraint in Section \ref{sec:obs}, where we also discuss that this constraint is also compatible with other constraints which can be obtained by considering higher-derivative effects in tidal Love numbers or ringdown quasinormal modes.

\section{Higher-derivative gravity and rotating black holes}
\label{sec:HDG}
In this section, we first review the possible higher-derivative invariants in four dimensions to eight-derivative order, and then review the deformation of the Kerr black hole due to these higher-derivative corrections.

\subsection{Higher-derivative extensions of general relativity}

\subsubsection*{Effective field theory of gravity}
Following an effective field theory approach, we may assume that the Lagrangian for our theory of gravity can be expanded in powers of derivatives. The two-derivative term corresponds to the Einstein-Hilbert Lagrangian, and, if diffeomorphism invariance is preserved, the rest of terms must be monomials formed out of contractions of the Riemann tensor $R_{\mu\nu\rho\sigma}$ and its covariant derivatives. The action that we will use in this paper takes the form \cite{Endlich:2017tqa,Cano:2019ore}

\begin{equation}\label{eq:EFTofGR}
S=\frac{1}{16\pi G}\int d^4x\sqrt{-g}\left\{R+\ell^4\mathcal{L}_{(6)}+\ell^6\mathcal{L}_{(8)}+\ldots \right\}\, ,
\end{equation}
where $\ell$ is a certain length scale, the six- and eight-derivative Lagrangians read
\begin{align}
\mathcal{L}_{(6)} &= \lambda_{\rm ev}\tensor{R}{_{\mu\nu }^{\rho\sigma}}\tensor{R}{_{\rho\sigma }^{\delta\gamma}}\tensor{R}{_{\delta\gamma }^{\mu\nu }}+\lambda_{\rm odd}\tensor{R}{_{\mu\nu}^{\rho\sigma}}\tensor{R}{_{\rho\sigma }^{\delta\gamma }} \tensor{\tilde R}{_{\delta\gamma }^{\mu\nu }} \, ,\\
\mathcal{L}_{(8)}&=\epsilon_1\mathcal{C}^2+\epsilon_2\tilde{\mathcal{C}}^2+\epsilon_3\mathcal{C}\tilde{\mathcal{C}}\, ,
\end{align}
where
\begin{equation}
\mathcal{C}=R_{\mu\nu\rho\sigma} R^{\mu\nu\rho\sigma}\, ,\quad \tilde{\mathcal{C}}=R_{\mu\nu\rho\sigma} \tilde{R}^{\mu\nu\rho\sigma}\, ,
\end{equation}
and
\begin{equation}
{\tilde R}^{\mu\nu\rho\sigma}=\frac{1}{2}\epsilon^{\mu\nu\alpha\beta}\tensor{R}{_{\alpha\beta}^{\rho\sigma}}
\end{equation}
is the dual of the Riemann tensor. 
Although the couplings in the EFT \req{eq:EFTofGR} are in principle free, they can be constrained by demanding that the theory satisfies reasonable physical conditions, such as causality \cite{Gruzinov:2006ie,Endlich:2017tqa,deRham:2020zyh,deRham:2021bll,Serra:2022pzl}, which imposes some constraints on the signs of the couplings (e.g. $\epsilon_{1,2}>0$ and $\epsilon_3^2\leq \epsilon_1\epsilon_2$).

In deriving this action, one makes use of various identities that reduce the number of curvature invariants. In addition, redefinitions of the metric allow one to get rid of all the densities that contain Ricci curvature. The ones that are at least quadratic in the latter are indeed irrelevant because they do not modify Einstein's vacuum solutions, and the redefinition that removes them is vanishing or of higher order on-shell. 
On the other hand, the redefinition that cancels the terms linear in the Ricci curvature is non-vanishing on-shell. Taking this into account means that if we want to employ the simple action \req{eq:EFTofGR}, containing only the pure Riemann terms, we must note that the metric $g_{\mu\nu}$ in our action is ambigious with respect to field redefinitions, namely:
\begin{equation}\label{eq:fieldredefs}
g_{\mu\nu}^{\rm amb}=g_{\mu\nu}+\ell^4 \Delta^{(4)}_{\mu\nu}+\ell^6 \Delta^{(6)}_{\mu\nu}+\ldots
\end{equation}
with tensors  $\Delta^{(n)}_{\mu\nu}$ built out of the curvature and the metric $g_{\mu\nu}$. We explore all the redefinitions possible for the action \req{eq:EFTofGR} in Appendix \ref{app:redefinitions}. Many physical observables are invariant under such redefinitions of the metric; for example, this is known to be the case for black hole thermodynamic quantities \cite{Jacobson:1993vj} or for the quasinormal mode frequencies which determine the gravitational wave emission during the ringdown. However, it is not so clear a priori how the definition of the multipolar structure behaves under field redefinitions. We show in appendix \ref{app:redefinitions} that the multipole moments are indeed invariant under such field redefinition ambiguities, and thus the theory \req{eq:EFTofGR} captures the most general modification of the black hole multipoles structure to eight-derivative corrections.

As an effective field theory, the action \req{eq:EFTofGR} can capture the corrections to GR coming from any UV theory that preserves diffeomorphism invariance and that does not introduce additional massless degrees of freedom. As a relevant example, let us take note of the case of string theory. While the 10-dimensional effective actions of superstring theories do not contain cubic-curvature terms\footnote{These can nonetheless arise in lower dimensions through e.g. flux compactifications or from the worldvolume actions of  D$p$-branes.}, they do contain the following quartic term \cite{Gross:1986iv} ---see also the more recent Refs.~\cite{Garousi:2020lof, Liu:2022bfg}---
\begin{equation}
\frac{\zeta(3)\alpha'^3}{8}R_4=\frac{\zeta(3)\alpha'^3}{8}\left(R_{\mu \nu \rho \sigma}R^{\alpha\nu \rho \beta}+\frac{1}{2}R_{\mu \sigma \nu \rho}R^{\alpha \beta\nu \rho}\right)R\indices{^\mu^\tau_\epsilon_\alpha}R\indices{_\beta_\tau^\epsilon^\sigma} \in \mathcal{L}_{\rm ST}\, ,
\end{equation}
up to field-redefinition-ambiguous terms involving the Ricci curvature. This is in fact the leading higher-curvature term in the case of type IIB string theory. The effective action of heterotic string theory contains quadratic curvature operators \cite{Bergshoeff:1989de}, which dominate over $R_4$, but they are coupled to scalar fields and therefore do not enter into the framework of \req{eq:EFTofGR}, which assumes that the only relevant degree of freedom is the metric. We study an appropriate model of such a form further below. 

In four dimensions we should be able to relate the density $R_4$ to a combination of the quartic densities in \req{eq:EFTofGR}. It turns out that
\begin{equation}\label{eq:R4rel}
R_4\Big|_{D=4}=\frac{1}{32}\mathcal{C}^2+\frac{1}{32}\tilde{\mathcal{C}}^2\, ,
\end{equation}
again modulo Ricci curvature.  Thus, assuming a toroidal compactification, the type IIB string theory prediction for the leading correction to GR corresponds  to 
\begin{equation}\label{eq:STcorr}
\ell^6\epsilon_1^{\rm ST}=\ell^6\epsilon_2^{\rm ST}=\frac{\zeta(3)\alpha'^3}{256}, \quad \epsilon_3^{\rm ST}=0\, .
\end{equation}

The equations of motion of \req{eq:EFTofGR} can be expressed as having an effective energy-momentum tensor in the right-hand-side of Einstein's equations, as follows
\begin{equation}\label{EinsteinEqTeff}
G_{\mu \nu}=T^{\rm eff}_{\mu\nu}\, ,
\end{equation}
with
\begin{equation}
T^{\rm eff}_{\mu\nu}=\ell^4T^{(6)}_{\mu\nu}+\ell^6T^{(8)}_{\mu\nu}\, ,
\end{equation}
where
\begin{equation}\label{Tmunun}
T^{(n)}_{\mu\nu}=-\tensor{P}{^{(n)}_{(\mu}^{\rho \sigma \gamma}} \tensor{R}{_{\nu) \rho \sigma \gamma}} +\frac{1}{2}g_{\mu \nu} \mathcal{L}_{(n)}-2 \nabla^\sigma \nabla^\rho P^{(n)}_{(\mu| \sigma|\nu)\rho}\, ,
\end{equation}
and the tensor $P^{(n)}_{\mu\nu\rho \sigma}$ is defined as the partial derivative of the corresponding Lagrangian with respect to the Riemann tensor, which yields
\begin{align}
\label{eq:Pcubic}
P^{(6)}_{\mu\nu\rho \sigma}&=3\lambda_{\rm ev} \tensor{R}{_{\mu\rho}^{\alpha \beta}}\tensor{R}{_{\alpha\beta\rho\sigma}}+\frac{3\lambda_{\rm odd}}{2}\left(\tensor{R}{_{\mu\rho}^{\alpha \beta}}\tensor{\tilde R}{_{\alpha\beta\rho\sigma}}+\tensor{R}{_{\mu\rho}^{\alpha \beta}}\tensor{\tilde R}{_{\rho\sigma\alpha\beta}}\right)\, ,\\
\label{eq:Pquartic}
P^{(8)}_{\mu\nu\rho \sigma}&=4\epsilon_1 \mathcal{C} R_{\mu\nu\rho \sigma}+2 \epsilon_2\mathcal{\tilde C}\left(\tilde R_{\mu\nu\rho \sigma}+\tilde R_{\rho \sigma\mu\nu}\right)+\epsilon_{3}\left[2\mathcal{\tilde C}R_{\mu\nu\rho \sigma}+\mathcal{C}\left(\tilde R_{\mu\nu\rho \sigma}+\tilde R_{\rho \sigma\mu\nu}\right)\right]\, .
\end{align}
Note that, due to the last term in \req{Tmunun} the equations of motion are of fourth order. However, this is not an issue since we deal with these theories in a perturbative fashion: we start with a solution of Einstein's equations $G_{\mu\nu}=0$ and use this solution to evaluate $T_{\mu\nu}^{\rm eff}$. Then we feed this back in \req{EinsteinEqTeff} and now we have to solve again the Einstein's equations with a ``matter source'' which accounts for the $\mathcal{O}(\ell^4)$  or $\mathcal{O}(\ell^6)$ corrections to the metric.

\subsubsection*{Quadratic theories with scalars}
In the EFT \req{eq:EFTofGR} we have neglected all quadratic curvature terms because they do not modify the solutions of vacuum Einstein equations. However, they do introduce modifications when coupled to scalar fields. We could consider a general action of the form
\begin{equation}\label{quadraticgen}
\begin{aligned}
S=\frac{1}{16\pi G}\int d^4x\sqrt{-g}\bigg\{&R-\frac{1}{2}\Sigma_{AB}(\phi)\partial_{\mu}\phi^A\partial^{\mu}\phi^B+V(\phi)+f_1(\phi) \ell^2 R^2+f_2(\phi) \ell^2 R_{\mu\nu}R^{\mu\nu}\\&+f_3(\phi) \ell^2\mathcal{X}_4 
+f_4(\phi) \ell^2\nabla^2 R+f_5(\phi)\ell^2 R_{\mu\nu\rho\sigma} \tilde{R}^{\mu\nu\rho\sigma}\bigg\}\, ,
\end{aligned}
\end{equation}
where we have a non-linear sigma model for an arbitrary number of scalar fields, with a potential $V(\phi)$, and which couple to all of the 4-derivative densities through arbitrary (differentiable) functions $f_i(\phi)$. Here, $\mathcal{X}_4$ is the Gauss-Bonnet density, given by
\begin{equation}
\mathcal{X}_4=R_{\mu\nu\rho\sigma} R^{\mu\nu\rho\sigma}-4 R_{\mu\nu}R^{\mu\nu}+R^2\, .
\end{equation}
However, if one is only interested in the leading-order corrections to the vacuum GR solutions, a few simplifications apply: (i) the four-derivative terms with Ricci curvature are again irrelevant or can be removed via field redefinitions, and (ii) the scalars acquire non-vanishing values of order $\ell^2$. By assuming that the scalar fields are massless, it can then be seen \cite{Cano:2019ore} that the leading correction to vacuum GR metrics in any theory of the form \req{quadraticgen} is captured by the much simpler model

\begin{equation}\label{quadraticaction}
\begin{aligned}
S=\frac{1}{16\pi G}\int d^4x\sqrt{-g}\bigg\{&R-\frac{1}{2}(\partial \phi_1)^2-\frac{1}{2}(\partial \phi_2)^2+\alpha_1 \ell^2 \phi_1\mathcal{X}_4 \\
&+\alpha_2 \ell^2 \left(\phi_2 \cos\xi+\phi_1 \sin\xi\right) R_{\mu\nu\rho\sigma} \tilde{R}^{\mu\nu\rho\sigma}\bigg\}\, ,
\end{aligned}
\end{equation}
that contains two scalars and depends only on three parameters, $\alpha_1$, $\alpha_2$ and $\xi$. 
This action reduces to Einstein-dilaton-Gauss-Bonnet gravity \cite{Kanti:1995vq} for $\alpha_2=0$, dynamical Chern-Simons gravity \cite{Alexander:2009tp} for $\alpha_1=0$, $\xi=0$ and the effective action of heterotic string theory compactified on a six-torus \cite{Cano:2021rey} for $\alpha_1\ell^2=\alpha_2\ell^2=-\alpha'/8$ and $\xi=0$. We also note that this theory breaks parity if and only if $\alpha_1\alpha_2\sin\xi\neq 0$. For $\sin\xi=0$, $\phi_2$ becomes a pseudoscalar so the theory preserves parity. 

The effective energy momentum tensor entering in the right hand side of Einstein's equations \req{EinsteinEqTeff} in this case reads
\begin{equation}\label{Teffscalars}
\begin{aligned}
T^{\rm eff}_{\mu\nu}=&-\alpha_1 \ell^2g_{\nu\lambda}\delta^{\lambda \sigma \alpha\beta}_{\mu\rho\gamma\delta}  R^{\gamma\delta}{}_{\alpha\beta} \nabla^\rho\nabla_\sigma \phi_1+4\alpha_2 \ell^2  \nabla^\rho\nabla^\sigma\left[\tilde R_{\rho(\mu\nu)\sigma}\,\left(\phi_2\cos \xi +\phi_1\sin \xi \right)\right] \\
&+\frac{1}{2}\left[\partial_\mu \phi_1 \partial_\nu \phi_1-\frac{1}{2}g_{\mu\nu} \left(\partial \phi_1\right)^2\right]+\frac{1}{2}\left[\partial_\mu \phi_2 \partial_\nu \phi_2-\frac{1}{2}g_{\mu\nu} \left(\partial \phi_2\right)^2\right] \ .
\end{aligned}
\end{equation}
Note that the contribution of the Gauss-Bonnet density is of second order in derivatives, while the Pontryagin density actually yields third-order equations, because $\nabla^\rho\tilde R_{\rho\mu\nu\sigma}=0$. Since they avoid fourth-order equations, these theories could even be studied non-perturbatively in $\ell$,\footnote{This is at least the case for the EdGB theory, which allows for a well-posed initial value problem \cite{Kovacs:2020ywu}. However, this is probably not the case for dCS theory \cite{Delsate:2014hba}.} but we will nevertheless restrict to considering solutions perturbative in $\ell^2$. On the other hand, the equations of motion for the scalar field read
\begin{equation}
\begin{aligned}
\label{scalarEOM}
 \nabla^2 \phi_1&=-\alpha_1 \ell^2 \mathcal{X}_4 -\alpha_2 \ell^2 \sin\xi R_{\mu\nu\rho\sigma}\tilde R^{\mu\nu\rho\sigma}\ , \\
\nabla^2 \phi_2&=-\alpha_2 \ell^2   \cos \xi \mathcal{X}_4\, ,
\end{aligned}
\end{equation}
and they typically imply that the scalars acquire a non-trivial profile when the curvature is non-vanishing.

\subsection{Rotating black hole solutions}

An appropriate ansatz to parametrize deviations to the Kerr geometry is given by the following metric \cite{Cano:2019ore}
\begin{equation}\label{rotatingmetric0}
\begin{aligned}
ds^2=&-\left(1-\frac{2 M \rho}{\Sigma}-H_1\right)dt^2-\left(1+H_2\right)\frac{4 M a \rho \sin^2\theta}{\Sigma}dtd\phi+\left(1+H_3\right)\left(\frac{\Sigma}{\Delta}d\rho^2+\Sigma d\theta^2\right)\\
&+\left(1+H_4\right)\left(\rho^2+a^2+\frac{2 M  \rho a^2\sin^2\theta}{\Sigma}\right)\sin^2\theta d\phi^2\, ,
\end{aligned}
\end{equation}

\noindent
where 
\begin{equation}
\Sigma=\rho^2+a^2\cos^2\theta\, ,\quad \Delta=\rho^2-2M\rho+a^2\, .
\end{equation} 
and $H_{1,2,3,4}$ are functions of $\cos\theta$ and $\rho$ only. This ansatz fixes most of the gauge freedom associated to infinitesimal coordinate transformations of $(\rho,\theta)$, and it has the advantage that it forces the horizon to be located at the largest root of $\Delta=0$. In order to preserve asymptotic flatness and to ensure that $M$ represents the mass and $J=aM$ the angular momentum of the black hole, these functions must satisfy the following boundary conditions

\begin{equation}
H_1\Big|_{\rho\rightarrow\infty}=0\, ,\quad
2H_2+H_3\Big|_{\rho\rightarrow\infty}=0\, ,\quad
2M H_3-\rho^2\frac{\partial H_3}{\partial\rho}\Big|_{\rho\rightarrow\infty}=0\, ,\quad
H_3-H_4\Big|_{\rho\rightarrow\infty}=0\, .
\end{equation}
We note that $H_3$ (and therefore $H_2$ and $H_4$) can tend to a non-zero constant value at infinity. 
These conditions fix the solution up to residual gauge freedom of the ansatz \req{rotatingmetric0}. Unfortunately, the full solution cannot be obtained analytically. A simple way to go around this problem is to consider a series expansion in the dimensionless spin $\chi=a/M$. In that case, the relevant solution takes the form
\begin{equation}
H_{i}(\rho,\theta)=\sum_{n=0}^{\infty}\chi^n\sum_{k=0}^{k_{\rm max}(n)}\sum_{p=0}^{n}H_{i}^{(n,k,p)}\frac{\cos^{p}\theta}{\rho^k}\, ,
\end{equation} 
for constant coefficients $H_{i}^{(n,k,p)}$, so that each $\chi^n$ term is a polynomial in $\cos\theta$ and in $1/\rho$. In the case of the quadratic theory \req{quadraticaction} one can also solve the equations for the scalar fields \req{scalarEOM} with a similar expansion,
\begin{equation}
\phi_{i}(\rho,\theta)=\sum_{n=0}^{\infty}\chi^n\sum_{k=0}^{\tilde k_{\rm max}(n)}\sum_{p=0}^{n}\phi_{i}^{(n,k,p)}\frac{\cos^{p}\theta}{\rho^k}\, .
\end{equation} 
This result then has to be used to evaluate the effective energy-momentum tensor in Einstein's equations \req{EinsteinEqTeff}. The solutions were computed in \cite{Cano:2019ore, Cano:2020cao}, where \texttt{Mathematica} codes were provided to obtain the solutions at any given order in the spin --- see the ancillary files of those references in arXiv. 

In the context of this work it would actually be more interesting to express the solution as an asymptotic expansion in $1/\rho$ instead of an expansion in $\chi$. However, it turns out that the $1/\rho$ series has a more complicated form. In fact, the coefficient of $1/\rho^k$ has contributions from arbitrary orders in $\chi^n$, and so these coefficients are not just polynomials in $\chi$ (nor in $\cos\theta$). Thus, we will make use of the $\chi$-expansion in order to obtain the multipoles. In certain cases, we will then be able to extract the exact functional dependence on $\chi$ from this expansion.

\section{Defining multipoles for higher-derivative Kerr}
\label{sec:multipoleexp}
Three ways have been proposed to define gravitational multipoles for four-dimensional, asymptotically flat, stationary, vacuum spacetimes. The first was that of Geroch \cite{Geroch:1970cd} and Hansen \cite{Hansen:1974zz}. They use the timelike Killing vector $\xi$ (which is the unique timelike vector normalized at infinity to $\xi^2=-1$) to construct two scalar fields. The first is simply $\lambda = \xi^2$. For the second, we define the twist of $\xi$ as:
\be \omega_\mu = \epsilon_{\mu\nu\rho\sigma} \xi^\nu \nabla^\rho \xi^\sigma.\ee
Using that $\xi$ is a Killing vector, this one-form satisfies:
\be \partial_{[\mu} \omega_{\nu]} = -\epsilon_{\mu\nu\rho\sigma}\xi^\rho \tensor{R}{^\sigma _{\lambda}} \xi^\lambda,\ee
which vanishes for vacuum spacetimes with $R_{\mu\nu}=0$. The vanishing of this curl means a scalar $\omega$ must exist such that:
\be \omega_\mu = \partial_\mu \omega.\ee
$\omega$ provides the second scalar in the Geroch-Hansen formalism. One can then conformally compactify the spacetime, and the expansion of the scalars $\lambda$ and $\omega$ around the (compactified) point at infinity gives the two families of gravitational multipoles --- the mass multipoles $M_\ell$ and the current multipoles $S_\ell$.

While an elegant and manifestly coordinate-independent formalism, the Geroch-Hansen formalism is not always practical to execute. By contrast, Thorne developed a formalism to define and compute gravitational multipoles using ACMC (asymptotically Cartesian and mass-centered) coordinates \cite{Thorne:1980ru} --- one simply needs to find a coordinate system in which the metric satisfies the ACMC condition, and then the multipoles can be read off from the $1/r$ asymptotic expansion of the metric components; we discuss this formalism in more detail in Section \ref{sec:ACMCdecomposition}. The equivalence of the Geroch-Hansen and Thorne multipole definitions, for vacuum spacetimes, was proved by G\"ursel \cite{1983GReGr..15..737G}.

The third and most recent framework to define the multipole moments was given in \cite{Compere:2017wrj}. There, a family of \emph{multipole symmetries} were introduced, allowing for an application of the covariant phase-space formalism \cite{Iyer:1994ys, Barnich:2001jy, Barnich:2007bf, Regge:1974zd} in order to calculate the corresponding asymptotic Noether charges --- these are precisely the gravitational multipoles of the spacetime. (We discuss this in more detail in Section \ref{sec:Noethercharge}.) It was also shown in \cite{Compere:2017wrj} that for vacuum, stationary spacetimes, this formalism gives equivalent results as the Geroch-Hansen or Thorne methods.

It was shown in \cite{Mayerson:2022ekj} that the Geroch-Hansen formalism can be extended to arbitrary non-vacuum spacetimes, including higher-derivative-deformed spacetimes. Moreover, if such a spacetime admits a coordinate system which satisfies the ACMC condition, then the G\"ursel proof can be generalized to show that the Geroch-Hansen and Thorne ACMC formalisms still give equivalent multipoles. Since we are able to find an ACMC expansion for the higher-derivative-deformed Kerr solution (see Section \ref{sec:ACMCdecomposition}), we can conclude that the Geroch-Hansen \emph{and} ACMC formalisms are well-defined and equivalent for our higher-derivative black holes.

However, the method of \cite{Compere:2017wrj}, and in particular its equivalence with Geroch-Hansen (or Thorne's ACMC) method, has not been considered beyond vacuum solutions in two-derivative gravity. In Section \ref{sec:Noethercharge}, we consider the generalization of the covariant phase-space formalism for multipoles in the presence of higher-derivative corrections, and show that it remains equivalent to the other two formalisms.

A final subtlety in the definition of gravitational multipoles is that of field redefinitions, i.e. $g_{\mu\nu}\rightarrow g_{\mu\nu} + X_{\mu\nu}$, where $X_{\mu\nu}$ is a tensor constructed from curvature tensors as in (\ref{eq:fieldredefs}). At first sight, such redefinitions can be concerning --- especially in Thorne's ACMC formalism, where multipoles are read off from a $1/r$ expansion in appropriate (ACMC) coordinates, and one may worry whether these field redefinitions could shift the value of the multipoles. Fortunately, we show in Appendix \ref{app:redefinitions} that any such field redefinitions will \emph{not} change the gravitational multipoles for our higher-derivative Kerr spacetime.

\subsection{ACMC decomposition}\label{sec:ACMCdecomposition}

For asymptotically flat, vacuum, four-dimensional spacetimes, a formalism for defining and calculating the gravitational multipoles of a spacetime was introduced by Thorne \cite{Thorne:1980ru}.

Thorne introduces the concept of an ACMC coordinate system, which allows one to read off all the mass and spin multipole moments from the expansion of the metric components at infinity. In practice, we will start from an AC system \cite{Bena:2020uup}\footnote{We will only work with axisymmetric spacetimes, thus presenting the ACMC formalism restricted to this case. The generalisation to non-axisymmetric spacetime can be found, for example, in the Appendix to \cite{Bena:2020uup}.}, which is not mass-centered and hence its mass dipole moment, $\tilde{M}_1$, might not vanish. Going to ACMC coordinates is then straightforward by shifting the origin of spacetime.

The AC coordinate system is defined by the following asymptotic expansion of the metric \cite{Bena:2020uup}:
 \begin{equation} \label{eq:theACMC}
\begin{aligned}
  g_{tt}  &= -1 + \frac{2M}{r}+ \sum_{\ell\geq 1}^{\infty}  \frac{2}{r^{\ell+1}} \left( \tilde M_\ell P_\ell + \sum_{\ell'<\ell} c^{(tt)}_{\ell \ell'} P_{\ell'} \right), \\
  g_{t\phi} &= -2r \sin^2\theta\left[ \sum_{\ell\geq 1}^{\infty} \frac{1}{r^{\ell+1}} \left( \frac{\tilde S_\ell}{\ell} P'_\ell +\sum_{\ell'<\ell} c_{\ell \ell'}^{(t\phi)}  P'_{\ell'}\right) \right],\\
  g_{rr} &= 1 + \sum_{\ell\geq 0}^\infty\frac{1}{r^{\ell+1}} \sum_{\ell'\leq \ell} c_{\ell \ell'}^{(rr)} P_{\ell'},\quad\quad
g_{\theta\theta} = r^2\left[ 1 + \sum_{\ell\geq 0}^\infty\frac{1}{r^{\ell+1}} \sum_{\ell'\leq \ell} c_{\ell \ell'}^{(\theta\theta)} P_{\ell'}\right],\\ 
g_{\phi\phi} &= r^2\sin^2\theta\left[ 1 + \sum_{\ell\geq 0}^\infty\frac{1}{r^{\ell+1}} \sum_{\ell'\leq \ell} c_{\ell \ell'}^{(\phi\phi)} P_{\ell'} \right],\quad
g_{r\theta} = (-r\sin\theta) \left[ \sum_{\ell\geq 0}^\infty\frac{1}{r^{\ell+1}} \sum_{\ell'\leq \ell} c_{\ell \ell'}^{(r\theta)} P'_{\ell'}\right],
\end{aligned}\end{equation}
where $P_l$ represents a Legendre Polynomial. The argument of the $P_l$ (and of their derivatives) in the expression above is always $\cos\theta$. The terms with coefficients $c^{(ij)}_{\ell\ell'}$ correspond to non-physical ``harmonics'', and depend on the particular AC(MC) system used. Even though these coefficients, $c^{(ij)}_{\ell\ell'}$, themselves are unphysical, the non-trivial condition for the expansion above to be AC is that all $c^{(ij)}_{\ell\ell'}$ have $\ell'\leq\ell$.

As already noted, the AC coordinate system above can be made into an ACMC one via a simple shift of the origin such that $\tilde{M}_1=0$ (this does not interfere with the condition $\ell'\leq \ell$ on the $c^{(ij)}_{\ell\ell'}$ coefficients). The gravitational multipoles can then be identified in the ACMC coordinate system as $M_\ell=\tilde{M}_\ell$ and $S_\ell=\tilde{S}_\ell$. In fact, there are simple formulae relating the true multipoles $M_\ell$ and $S_\ell$ in terms of the coefficients $\tilde{M}_\ell$ and $\tilde{S}_\ell$ in an arbitrary AC coordinate system, namely \cite{Bena:2020uup}: 
\begin{equation}
M_\ell=\sum_{k=0}^{\ell}\binom{\ell}{k}\tilde{M}_k\bigg(-\frac{\tilde{M}_1}{\tilde{M}_0}\bigg)^{\ell-k},\quad S_\ell=\sum_{k=0}^{\ell}\binom{\ell}{k}\tilde{S}_k\bigg(-\frac{\tilde{M}_1}{\tilde{M}_0}\bigg)^{\ell-k}.
\end{equation}

Strictly speaking, the discussion above concerns an ACMC-$\infty$ system. Were the expansions presented in (\ref{eq:theACMC}) only valid up to some finite order $N$, then we would have an ACMC-$N$ (or AC-$N$) coordinate system from which only the first $N+1$ multipoles can be read off \cite{Thorne:1980ru,Bena:2020uup}.

\subsubsection*{Application to higher-derivative Kerr}
The metric \req{rotatingmetric0} is not in ACMC form so we have to perform a change of coordinates to write it in that form. In the absence of higher-derivative corrections, \req{rotatingmetric0} corresponds to the Kerr metric in Boyer-Lindquist coordinates. A coordinate transformation to an ACMC-$\infty$ coordinate system $(\rho_{S},\theta_{S})$ is given by: 
\begin{equation}\label{eq:spheroidaltospherical}
\rho_{S}\sin\theta_{S}=\sqrt{\rho^2+a^2}\sin\theta\, ,\quad \rho_{S}\cos\theta_{S}=\rho\cos\theta\, ,
\end{equation}
where the ACMC coordinates $(\rho_S,\theta_S)$ can be thought of as asymptotically spherical coordinates (as opposed to Boyer-Lindquist coordinates $(r,\theta)$, which are (asymptotically) spheroidal).

It will be useful to introduce the notation $x=\cos\theta$, $x_{S}=\cos\theta_{S}$. The relations (\ref{eq:spheroidaltospherical}) can be solved explicitly to obtain the change of coordinates 
\begin{equation}\label{eq:rhox0}
\begin{aligned}
\rho&\equiv \rho^{(0)}(\rho_{S},x_{S})=\frac{\sqrt{\rho_{S}^2-a^2+\sqrt{\rho_{S}^4+a^4+2 a^2 \rho_{S}^2  \left(2 x_{S}^2-1\right)}}}{\sqrt{2}}\, ,\\
x&\equiv x^{(0)}(\rho_{S},x_{S})=\frac{\rho^{(0)}(\rho_{S},x_{S})}{2a^2\rho_{S}x_{S}}\left(-\rho_{S}^2+a^2+\sqrt{\rho_{S}^4+a^4+2 a^2 \rho_{S}^2  \left(2 x_{S}^2-1\right)}\right)\, .
\end{aligned}
\end{equation}
However, this does not put the metric into the ACMC form when higher-derivative corrections are involved. Thus, we need to search for a more general transformation. Let us denote by $\mu$ the parameter that controls the leading corrections, so that $\mu=\ell^4$ or $\mu=\ell^6$ depending on the case. At linear order in $\mu$ we then must consider a change of coordinates of the form

\begin{equation}
\begin{aligned}
\rho&=\rho^{(0)}(\rho_{S},x_{S})+\mu\, \rho^{(1)}(\rho_{S},x_{S})+\ldots\, ,\\ x&=x^{(0)}(\rho_{S},x_{S})+\mu\, x^{(1)}(\rho_{S},x_{S})+\ldots
\end{aligned}
\end{equation}
where $\rho^{(0)}$ and $x^{(0)}$ are given above in (\ref{eq:rhox0}), and we have to find appropriate $\rho^{(1)}$ and $x^{(1)}$. Since we do not have a fully analytic form of the metric, we cannot obtain a closed expression for this change of coordinates. Nevertheless, we can work order by order in the $\rho_S^{-1}$ expansion to compute the multipoles order by order. Indeed, we find that it suffices to consider a change of coordinates of the form
\begin{equation}\label{coordinatechange}
\begin{aligned}
\rho^{(1)}(\rho_{S},x_{S})&=b_{0,0}\,\rho_{S}+\sum_{k=0}^{\infty}\sum_{p=0}^{k+1}b_{k,p}\frac{x_{S}^{p}}{\rho_{S}^k}\, ,\\
x^{(1)}(\rho_{S},x_{S})&=(1-x_{S}^2)\sum_{k=1}^{\infty}\sum_{p=0}^{k-1}c_{k,p}\frac{x_{S}^{p}}{\rho_{S}^k}
\end{aligned}
\end{equation}
for certain coefficients $b_{k,p}$ and $c_{k,p}$. We highlight the $b_{0,0}$ term as this one is somewhat special; its presence is necessary as the coordinate $\rho$ in  \req{rotatingmetric0} does not have the usual asymptotic behavior. To see this, we note that the $H_i$ functions behave asymptotically as 
\begin{align}
H_1&=-\frac{h_3 M}{\rho}+\mathcal{O}\left(\frac{1}{\rho^2}\right)\,, \quad  H_2=-\frac{h_3}{2}+\mathcal{O}\left(\frac{1}{\rho}\right)\,, \\
H_3&=h_3\left(1-\frac{M}{\rho}\right)+\mathcal{O}\left(\frac{1}{\rho^2}\right)\,, \quad H_4=h_3+\mathcal{O}\left(\frac{1}{\rho}\right)\, .
\end{align}
for a certain constant $h_3$. Therefore, the metric reads
\begin{equation}
\begin{aligned}
ds^2(\rho\rightarrow\infty)=&-\left(1-\frac{2 M(1-h_3/2) }{\rho}\right)dt^2-(1-h_3/2)\frac{4 M a \sin^2\theta}{\rho}dtd\phi\\
&+(1+h_3)\left(d\rho^2+\rho^2d\theta^2
+\rho^2\sin^2\theta d\phi^2\right)\, ,
\end{aligned}
\end{equation}
so that the metric is not even ACMC-0. The metric can be made ACMC-0 if we redefine the radial coordinate by $\rho=(1-h_3/2)\rho_{S}+\mathcal{O}(\rho_{S}^0)$, so that we identify $\mu\,b_{0,0}=-h_3/2$. This allows us to read off the mass and angular momentum from the asymptotic expansion, $M_{0}=M$, $S_{1}=a M$. The rest of the coefficients in \req{coordinatechange} are determined similarly at increasing orders in $1/\rho$ by demanding that the asymptotic expansion of the metric contains no terms of the form  $x_{S}^m/\rho_{S}^n$, with $m\geq n$. This is equivalent to the statement that all $c^{(ij)}_{\ell\ell'}$ in \eqref{eq:theACMC} have $\ell'\leq\ell$. Once that is done, we can read off the rest of the multipoles, up to the order to which the aforementioned condition holds.

\subsection{Equivalence of the covariant phase space approach}\label{sec:Noethercharge}

Thorne's ACMC formalism of the previous section provides an easy way to obtain multipole moments in a certain (ACMC) asymptotic expansion of the metric. On the other hand, the elegant formalism of \cite{Compere:2017wrj} defines the physical multipole moments through surface integrals at infinity associated to the vector fields that generate the so-called \emph{multipole symmetries}\footnote{For our purposes here we do not need to show the explicit expressions of these vector fields. These can be found in Sec.~2 of \cite{Compere:2017wrj}, where it is explained that one can make use of \eqref{eq:defmultipoles} in order to compute the associated Noether charges.}. 
For stationary solutions in Einstein gravity, the authors of \cite{Compere:2017wrj} showed that their definition of mass and current multipole moments fully agrees with Thorne's ACMC definitions.
However, it is not clear that the surface charge multipole moments of \cite{Compere:2017wrj} will continue agreeing with Thorne's ACMC multipoles beyond vacuum GR solutions. Indeed, it is for example well-known that the (e.g. ADM) definitions of energy, mass and angular momentum are modified in theories with higher-derivative corrections  \cite{Deser:2002jk,Senturk:2012yi,Adami:2017phg}.

In higher-derivative theories, there can a priori be two classes of corrections to the black hole multipoles. The first are the corrections due to the perturbation of the metric itself --- these are clearly captured by the asymptotic expansion in the ACMC formalism (as we calculate below in Section \ref{sec:BHmultipoles}). On the other hand, there could also be corrections to the \emph{definitions} themselves of the multipole moments --- in other words, the functional form of surface charge multipole moments of \cite{Compere:2017wrj} may change as well. Such corrections would not be captured in the ACMC formalism, leading to possible mismatches between the surface charge and ACMC multipole moment definitions. The purpose of this section is to show that these corrections to the functional form of the surface charge non-trivially vanish for asymptotically flat black holes in our higher-derivative theories, so that the Thorne ACMC and surface charge multipole definitions remain equivalent for these theories.

Let us first consider the cubic and quartic theories. For these, the surface integrals that one has to compute in order to extract the multipole moments are \cite{Compere:2017wrj}:

\begin{equation}\label{eq:defmultipoles}
{\cal Q}_{\xi}[\delta g]=\lim_{r\to \infty}\left(\frac{1}{8\pi G}\int_{S^2_{\infty}}\textbf{k}_{\xi}[\delta g;\eta]\right)\,, 
\end{equation}
where $\textbf{k}_{\xi}[\delta g;\eta]$ can be taken to be the Iyer-Wald 2-form associated to the vector $\xi$ and to the metric perturbation ${\delta g}_{\mu\nu}$ over the background metric, which is taken to be the Minkowski metric, $\eta_{\mu\nu}$. The corrections to the definition of the multipoles are encoded in the Iyer-Wald 2-form, which is a theory-dependent quantity. For ${\cal L}\left(R_{\mu\nu\rho\sigma}, g^{\alpha\beta}\right)$ theories, there exists a well-known expression for the Iyer-Wald 2-form (see e.g. \cite{Bueno:2016ypa} and also Appendix~\ref{app:Waldformalism}), which is the following 
\begin{equation}\label{eq:IW2form}
\begin{aligned}
\textbf{k}_{\xi}[\delta g;g]=\,&{\boldsymbol\epsilon}_{\mu\nu}\left[-{\delta P}^{\mu\nu\rho}{}_{\sigma}\nabla_{\rho}\xi^{\sigma}-2\xi^{\rho}\delta\left(\nabla_{\sigma}P^{\mu\nu}{}_{\rho}{}^{\sigma}\right)\right.\\[1mm]
&\left.+\delta g_{\alpha\beta}\left(-\frac{1}{2}P^{\mu\nu\rho\sigma}\nabla_{\rho}\xi_{\sigma}g^{\alpha\beta}+2\xi^{\nu}\nabla_{\lambda}P^{\mu\alpha\beta\lambda}-\xi_{\rho}\nabla_{\sigma}P^{\mu\nu\rho\sigma}g^{\alpha\beta}\right)\right.\\[1mm]
&\left.-\nabla_{\lambda}\delta g_{\alpha\beta}\left(\xi^{\alpha}P^{\mu\nu\lambda\beta}+2\xi^{\nu}P^{\mu\alpha\beta\lambda}\right)\right]\, ,
\end{aligned}
\end{equation}
where 
\begin{equation}
{\boldsymbol\epsilon}_{\mu\nu}\equiv\frac{1}{2!}\epsilon_{\mu\nu\rho\sigma}dx^{\rho}\wedge dx^{\sigma}\, ,
\end{equation}
and the tensor $P^{\mu\nu\rho\sigma}$ is defined as
\begin{equation}
P^{\mu\nu\rho\sigma}=\frac{\partial {\cal L}}{\partial R_{\mu\nu\rho\sigma}}\, ,
\end{equation}
which is assumed to satisfy
\begin{equation}
P_{\mu\nu\rho\sigma}=-P_{\nu\mu\rho\sigma}\,, \hspace{1cm} P_{\mu\nu\rho\sigma}=P_{\rho\sigma\mu\nu}\,, \hspace{1cm} P_{\mu[\nu\rho\sigma]}=0\,.
\end{equation}
For the cubic and quartic theories, one gets\footnote{Note that we subtract the antisymmetric part $P^{(n)}_{\mu[\nu\rho\sigma]}$, since the tensors $P^{(6)}_{\mu\nu\rho\sigma}$ and $P^{(8)}_{\mu\nu\rho\sigma}$ as defined in Eqs.~\req{eq:Pcubic}, \req{eq:Pquartic} do not satisfy $P_{\mu[\nu\rho\sigma]}=0$. However, one can check that the term $P^{(n)}_{\mu[\nu\rho\sigma]}$ does not change the stress-energy tensor \req{Tmunun}, so we could have defined these tensors directly in this way.}
\begin{equation}
P_{\mu\nu\rho\sigma}=g_{\mu[\rho}g_{\sigma]\nu}+\ell^{n-2} \left(P^{(n)}_{\mu\nu\rho\sigma}-P^{(n)}_{\mu[\nu\rho\sigma]}\right)\, .
\end{equation}
From eqs.~\eqref{eq:Pcubic} and \eqref{eq:Pquartic}, we see that $P^{(6)}_{\mu\nu\rho\sigma}$ and $P^{(8)}_{\mu\nu\rho\sigma}$ are quadratic and cubic in the Riemann tensor, respectively.  Their variations $\delta P$ are then linear and quadratic in the background curvature. This implies that all the corrections to the Iyer-Wald 2-form (and therefore to the definition of the multipole moments) vanish when the background metric is flat,
\begin{equation}
\textbf{k}_{\xi}[\delta g;\eta]=\textbf{k}^{\rm {GR}}_{\xi}[\delta g; \eta]\,.
\end{equation}
The argument for the quadratic theories is basically the same, but it is convenient to treat them separately as they contain additional scalar degrees of freedom. In the latter case, the surface integrals that one has to compute are
\begin{equation}\label{eq:defmultipoles2}
{\cal Q}_{\xi}\left[\delta g, \delta \phi^{A}\right]=\lim_{r\to \infty}\left(\frac{1}{8\pi G}\int_{S^2_{\infty}}\textbf{k}_{\xi}\left[\delta g, \delta \phi^A;\eta, \phi^A_{\infty}\right]\right)\,,
\end{equation}
where $\phi^{A}_{\infty}$ correspond the asymptotic values of the scalars ($A=1, 2$). Since the quadratic theories are invariant (up to total derivatives) under constant shifts of the scalars, we can choose $\phi^{A}_{\infty}=0$ without loss of generality.  The expression of the Iyer-Wald 2-form for this class of theories is derived in Appendix~\ref{app:Waldformalism} to be:
\begin{equation}\label{eq:IW2form2}
\begin{aligned}
\textbf{k}_{\xi}\left[\delta g, \delta \phi^A ;g, \phi^A\right]=\,&{\boldsymbol\epsilon}_{\mu\nu}\left[-{\delta P}^{\mu\nu\rho}{}_{\sigma}\nabla_{\rho}\xi^{\sigma}-2\xi^{\rho}\delta\left(\nabla_{\sigma}P^{\mu\nu}{}_{\rho}{}^{\sigma}\right)\right.\\[1mm]
&\left.+\delta g_{\alpha\beta}\left(-\frac{1}{2}P^{\mu\nu\rho\sigma}\nabla_{\rho}\xi_{\sigma}g^{\alpha\beta}+2\xi^{\nu}\nabla_{\lambda}P^{\mu\alpha\beta\lambda}-\xi_{\rho}\nabla_{\sigma}P^{\mu\nu\rho\sigma}g^{\alpha\beta}\right)\right.\\[1mm]
&\left.-\nabla_{\lambda}\delta g_{\alpha\beta}\left(\xi^{\alpha}P^{\mu\nu\lambda\beta}+2\xi^{\nu}P^{\mu\alpha\beta\lambda}\right)+\xi^{\nu}\delta_{AB}\partial^{\mu}\phi^{B}\delta \phi^{A}\right]\, ,
\end{aligned}
\end{equation}
where 
\begin{equation}
P_{\mu\nu\rho\sigma}=g_{\mu[\rho}g_{\sigma]\nu}+\ell^{2} \left(P^{(4)}_{\mu\nu\rho\sigma}-P^{(4)}_{\mu[\nu\rho\sigma]}\right)\, ,
\end{equation}
and 
\begin{equation}
P^{(4)}_{\mu\nu\rho\sigma}=\,-2\alpha_{1}\phi_1 {\tilde{\tilde R}}_{\mu\nu\rho\sigma}+\alpha_2\left(\phi_2 \cos\xi+\phi_1\sin\xi\right)\left({\tilde R}_{\mu\nu\rho\sigma}+{\tilde R}_{\rho\sigma\mu\nu}\right)\,,
\end{equation}
where ${\tilde{\tilde R}}_{\mu\nu\rho\sigma}$ the double-dual of the Riemann tensor. As we can see, the expressions for the Iyer-Wald 2-form, eqs.~\eqref{eq:IW2form} and \eqref{eq:IW2form2}, are almost identical except for the last term in \eqref{eq:IW2form2}, which in any case vanishes for constant background scalars. One can further check that the remaining corrections to the Iyer-Wald 2-form also vanish since they contain either one Riemann tensor of the background metric or the background scalars, both of which vanish.

We can conclude that the multipole moments of asymptotically flat black holes in the theories \req{eq:EFTofGR} and \req{quadraticaction}
can be identified exactly as in GR.\footnote{An alternative way to arrive to the same conclusion is by noticing that the linearized theories around Minkowski (which according to \cite{Compere:2017wrj} is all we need to compute the multipoles) are the same as in GR. This will no longer be the case if we consider spacetimes with different asymptotics, for which multipole moments can also be defined, see e.g.~\cite{Mukherjee:2020how, Chakraborty:2021ezq}.} In particular, the multipole symmetry formalism of \cite{Compere:2017wrj} and Thorne's ACMC formalism \cite{Thorne:1980ru} will give equivalent definitions of the gravitational multipoles for higher-derivative-deformed Kerr.

\section{Black hole multipoles}
\label{sec:BHmultipoles}

In this section, we give the multipoles of the higher-derivative Kerr solution (\ref{rotatingmetric0}), which are calculated using the ACMC methods described above in Section \ref{sec:ACMCdecomposition}. We will discuss how the multipoles change from their two-derivative Kerr values separately for six-, eight- and four-derivative gravitational corrections. Note that additional explicit expressions of the multipoles (and expansions to higher orders in $\chi$) are available in the ancillary \texttt{Mathematica} notebook attached to this paper's arXiv version.
Finally, we will also discuss how ratios of these multipoles behave, and the relation to the conjecture on these ratios in \cite{Bena:2020see,Bena:2020uup}.

It is useful to define the dimensionless couplings
\begin{equation}\label{eq:dimlessallcouplings}
\hat\lambda_{\rm ev}=\frac{\ell^4}{M^4}\lambda_{\rm ev}\, ,\quad \hat\lambda_{\rm odd}=\frac{\ell^4}{M^4}\lambda_{\rm odd}\, ,\quad \hat\epsilon_i=\frac{\ell^6}{M^6}\epsilon_i\, , \quad 
\hat\alpha_{i}=\frac{\ell^2}{M^2}\alpha_i\, .
\end{equation}
and parametrize the multipoles as deviations from Kerr
\begin{equation}
M_\ell=M_\ell^{(0)}+\delta M_\ell,\quad S_\ell=S_\ell^{(0)}+\delta S_\ell,
\end{equation}
where $M_\ell^{(0)}$ and $S_\ell^{(0)}$ represent the Kerr multipoles, given by:
\begin{equation}
M_{2n}=M(-a^2)^n,\quad M_{2n+1}=0,\quad S_{2n}=0,\quad S_{2n+1}=M\,a(-a^2)^n.
\end{equation}

\subsection{Six-derivative corrections}\label{sec:mult6D}
Upon computation of the multipoles $M_{n}$, $S_{n}$ for a few values of $n$, we observe the following properties. On the one hand, the parity-preserving corrections $\hat\lambda_{\rm ev}$ only affect the even mass multipoles $M_{2n}$ and the odd current multipoles $S_{2n+1}$. On the other, the parity-breaking corrections  $\hat\lambda_{\rm odd}$ modify the multipoles $M_{2n+1}$ and $S_{2n}$, which are vanishing for Kerr. These odd-parity corrections and multipoles break the equatorial symmetry of Kerr. Furthermore, while for Kerr the multipole moments take the simple functional dependence $\sim M a^n$, we find that the corrections are in general complicated functions of the spin; they behave as $\sim \chi^n$ only for small $\chi$, whereas for $\chi\sim 1$ they tend to a constant value and the series expansion contains half-integer powers as well.
By analyzing the first 30 values of $n$, we observe an intriguing connection between the corrections associated to $\hat\lambda_{\rm ev}$ and $\hat\lambda_{\rm odd}$. We have 
\begin{gather}
\delta S_{2\,n}=\frac{\hat\lambda_{\rm odd}}{\hat\lambda_{\rm ev}}\delta M_{2\,n},\notag\\
\delta S_{2\,n+1}=-\frac{\hat\lambda_{\rm ev}}{\hat\lambda_{\rm odd}}\delta M_{2\,n+1}\, ,
\label{eq:sixDervRels1}
\end{gather}
and we recall that $\delta S_{2\,n+1}$ and $\delta M_{2\,n}$ are proportional to $\hat\lambda_{\rm ev}$ while $\delta S_{2\,n}$ and $\delta M_{2\,n+1}$ are proportional to $\hat\lambda_{\rm odd}$. These relations seem indeed to hold for arbitrary $n$ and $a$, and they allow us to express the mass and current multipole moments in a more compact form by introducing the complex multipoles
\begin{equation}
Z_n=M_n+i S_n\, .
\end{equation}
For Kerr, $Z_n$ takes the simple form $Z_n^{(0)}=M(ia)^n$, and due to (\ref{eq:sixDervRels1}), we also find a very appealing formula for the six-derivative theories 
\begin{equation}
Z_{n}=Z_n^{(0)}\left[1+\left(\hat\lambda_{\rm ev}+i \hat\lambda_{\rm odd}\right)f_n(\chi)\right]\, .
\label{eq:zn6Derv}
\end{equation}
Thus, the full dependence on the higher-order parameters is encoded only through the complex coupling constant $\hat\lambda_{\rm ev}+i \hat\lambda_{\rm odd}$. The functions $f_n(\chi)$ are dimensionless and they capture the relative deviation with respect to the Kerr multipoles. From the method explained in the previous section, we can obtain the series expansion of these functions around $\chi=0$. However, it turns out that, at least for low values of $n$, these expansions take a sufficiently simple structure that allows us with the help of \texttt{Mathematica}\footnote{One can use \textit{FindSequenceFunction} with enough terms (usually 20+) in the $\chi$ expansion of $f_n(\chi)$ to get a general expression for the coefficient of the expansion. That can be summed to obtain the given expressions.} to recognize the pattern and to find the corresponding generating function. Thus, we are able to obtain the exact result, which for the $n=2,3$ and $4$ multipoles reads
\begin{align}
\label{eq:6Df2full} f_{2}(\chi)&=-\frac{4}{7\,\chi^6}\big(8-4\chi^6+15\chi^4-20\chi^2-8(1-\chi^2)^{5/2}\big)\, ,\\
f_{3}(\chi)&=\frac{4}{7\,\chi^8}\bigg[4\sqrt{1-\chi^2}\,(8\chi^6-10\chi^4+\chi^2+16)-16+4\chi^8-13\chi^6+10\chi^4-8\chi^2\notag\\
&\hspace{1.5cm}+15\chi\arcsin(\chi)\bigg]\\\notag
\label{eq:6Df4full}f_{4}(\chi)&=\frac{8}{49\,\chi^8}\bigg[\sqrt{1-\chi^2}\,(56\chi^6-108\chi^4+48\chi^2-311)+28\chi^8-103\chi^6+136\chi^4\notag\\
&\hspace{1.5cm}+24\chi^2+416-\frac{840\,(1+2\chi^2)}{\chi}\arcsin(\chi)\bigg],
\end{align}
and, by construction, $f_0(\chi)=f_1(\chi)=0$.

Note that, despite the powers of $\chi$ appearing in the denominators, these functions take a finite value in the limit $\chi\rightarrow 0$. In fact, we have
\begin{align}
\label{eq:6Df2pert}f_{2}(\chi)&=\frac{6}{7}-\frac{5}{28}\chi^2-\frac{3}{56}\chi^4+\mathcal{O}\big(\chi^6\big)\, ,\\
f_{3}(\chi)&=\frac{69}{98}-\frac{19}{84}\chi^2-\frac{13}{176}\chi^4+\mathcal{O}\big(\chi^6\big)\, ,\\
\label{eq:6Df4pert}f_{4}(\chi)&=\frac{242}{147}-\frac{1241}{3234}\chi^2-\frac{842}{7007}\chi^4+\mathcal{O}\big(\chi^{6}\big)\, .
\end{align}

\begin{figure}[t!]
	\centering
	\includegraphics[width=\textwidth]{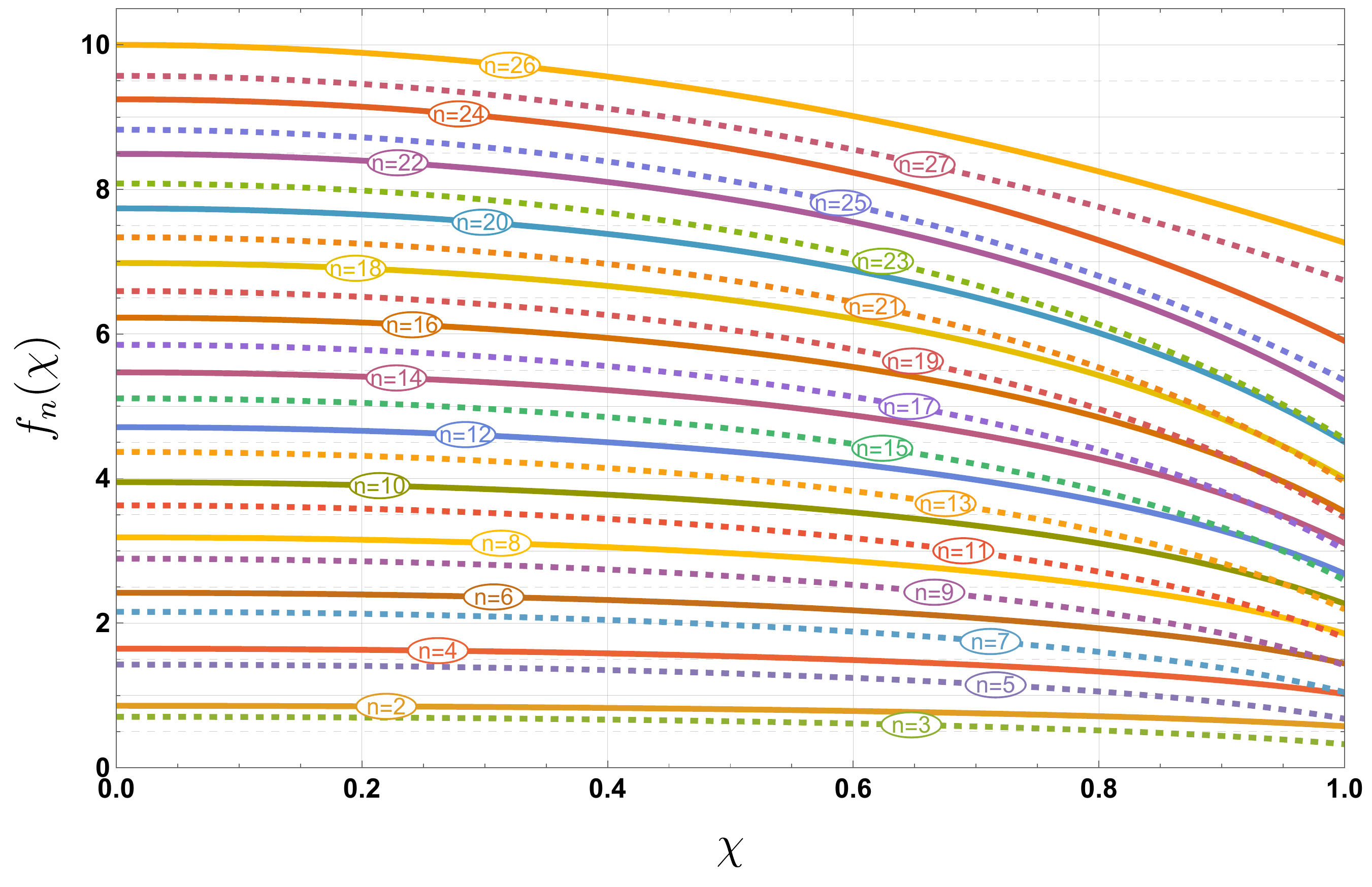}
	\caption{The function $f_n(\chi)$ determining the six-derivative corrections to the multipoles in \eqref{eq:zn6Derv}. Odd and even multipoles are given by dashed and solid lines, respectively.}
	\label{fig:sixDervMassCor}
\end{figure}

\begin{figure}[t!]
	\centering
	\includegraphics[width=\textwidth]{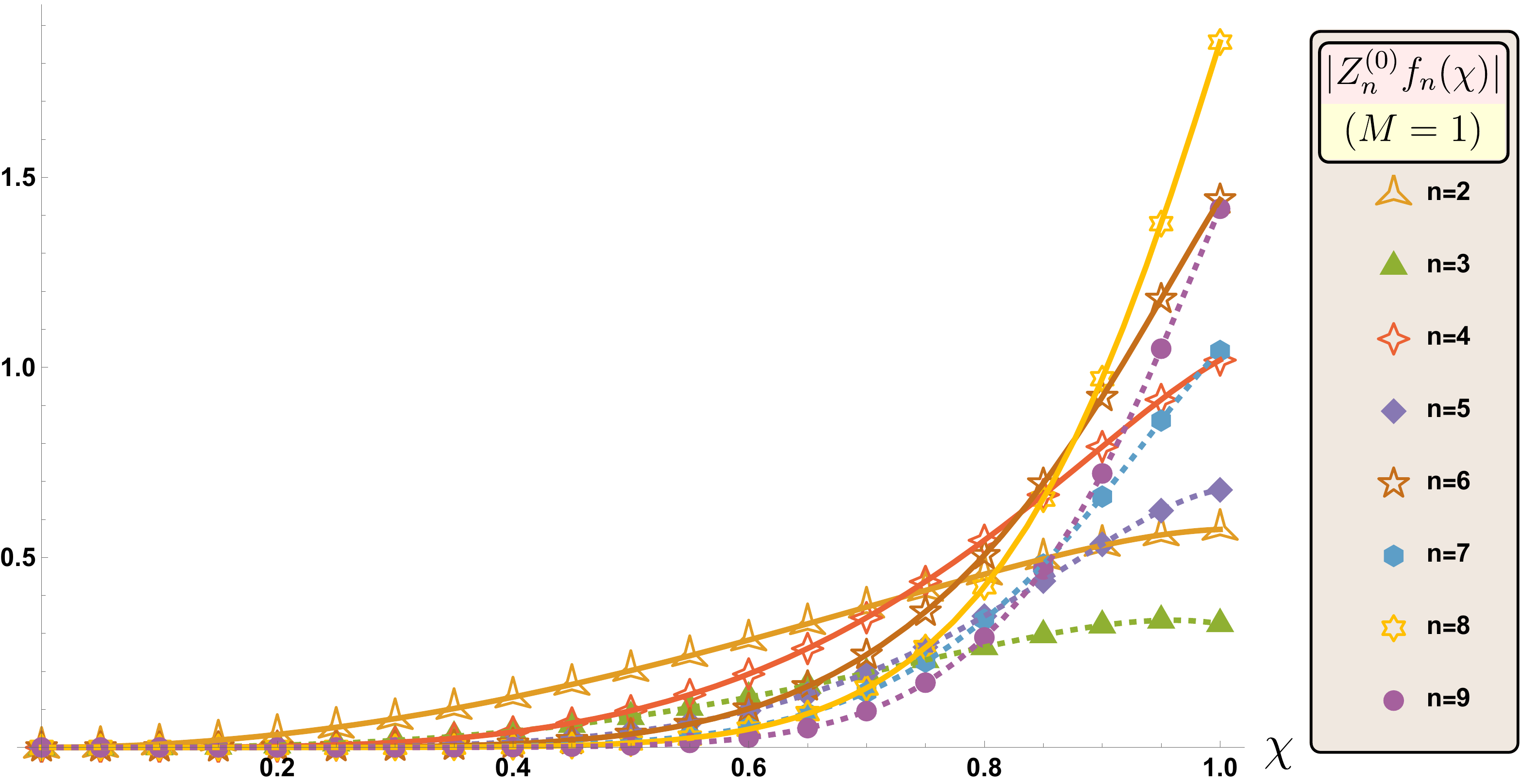}
	\caption{Absolute correction, $\lvert Z_n^{(0)}f_n(\chi)\rvert$, to the Kerr multipoles for six-derivative theories. Only the first 8 non-trivial terms are shown for clarity.}
	\label{fig:sixDervMassCor2}
\end{figure}

We plot the $f_n$ in Fig.~\ref{fig:sixDervMassCor} and we see that they are decreasing functions of the spin $\chi$, so the relative correction with respect to Kerr is larger for smaller spin. That is not the case for the absolute correction, $Z_n^{(0)}f_n(\chi)$, see Fig.~\ref{fig:sixDervMassCor2}. Nevertheless, the dependence on the spin is quite mild.

For $n=2,3,4$ the lines in Figure \ref{fig:sixDervMassCor} are plotted using the exact expressions. They are virtually indistinguishable from the same curves drawn with a series expansion to order $\chi^{20}$. We have calculated all the given multipoles to order $\chi^{30}$, hence one can argue that the first few curves can be trusted almost all the way to extremality. 

\begin{figure}[t!]
	\centering
	\includegraphics[width=\textwidth]{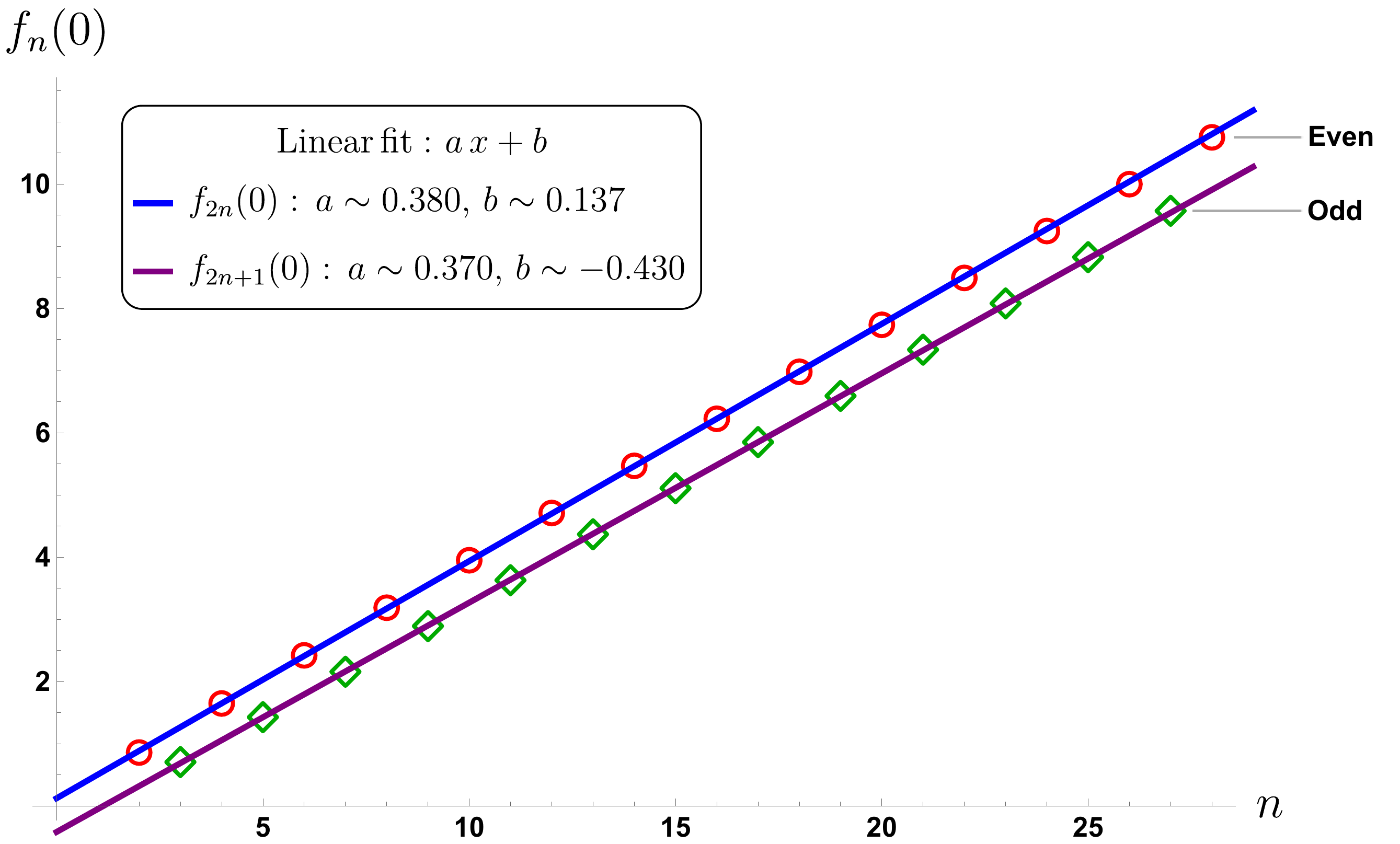}
	\caption{Linear fits to $f_n(0)$ as a function of $n$. Done separately for even and odd $n$.}
	\label{fig:sixDervLinGrowth}
\end{figure}

Regarding the dependence with $n$, we are not able to find a simple exact formula for the general term $f_n$ - not even for $f_n(0)$ - but from Fig.~\ref{fig:sixDervMassCor} we expect a linear growth with $n$, which is exhibited more explicitly in Fig.~\ref{fig:sixDervLinGrowth}. As we can see, the $f_n(0)$ have a slightly different behavior for odd and even $n$, but both series are almost perfectly linear with $n$. Interestingly, this implies that, no matter how small the coupling constants $\lambda_{\rm ev,odd}$ are, the black hole multipole moments will receive $\mathcal{O}(1)$ corrections when $n$ is large. According to the fits in Fig.~\ref{fig:sixDervLinGrowth} this will happen for 
\begin{equation}
n_{\rm max}\sim \frac{2}{\left|\hat\lambda_{\rm ev,odd}\right|}=\frac{2 M^4}{\ell^4\left|\lambda_{\rm ev,odd}\right|}\, ,
\end{equation}
and for $n>n_{\rm max}$ the effective field theory breaks down.

\subsection{Eight-derivative corrections}\label{sec:mult8D}
As in the six-derivative case, the even-parity corrections $\hat\epsilon_1$ and $\hat\epsilon_2$ modify the multipoles $M_{2n}$ and $S_{2n+1}$, while the parity-breaking term $\hat\epsilon_3$ gives rise to non-zero $M_{2n+1}$ and $S_{2n}$. The corrections are again non-polynomial functions of the spin which have non-trivial relations between them.

\begin{figure}[t!]
	\centering
	\includegraphics[width=\textwidth]{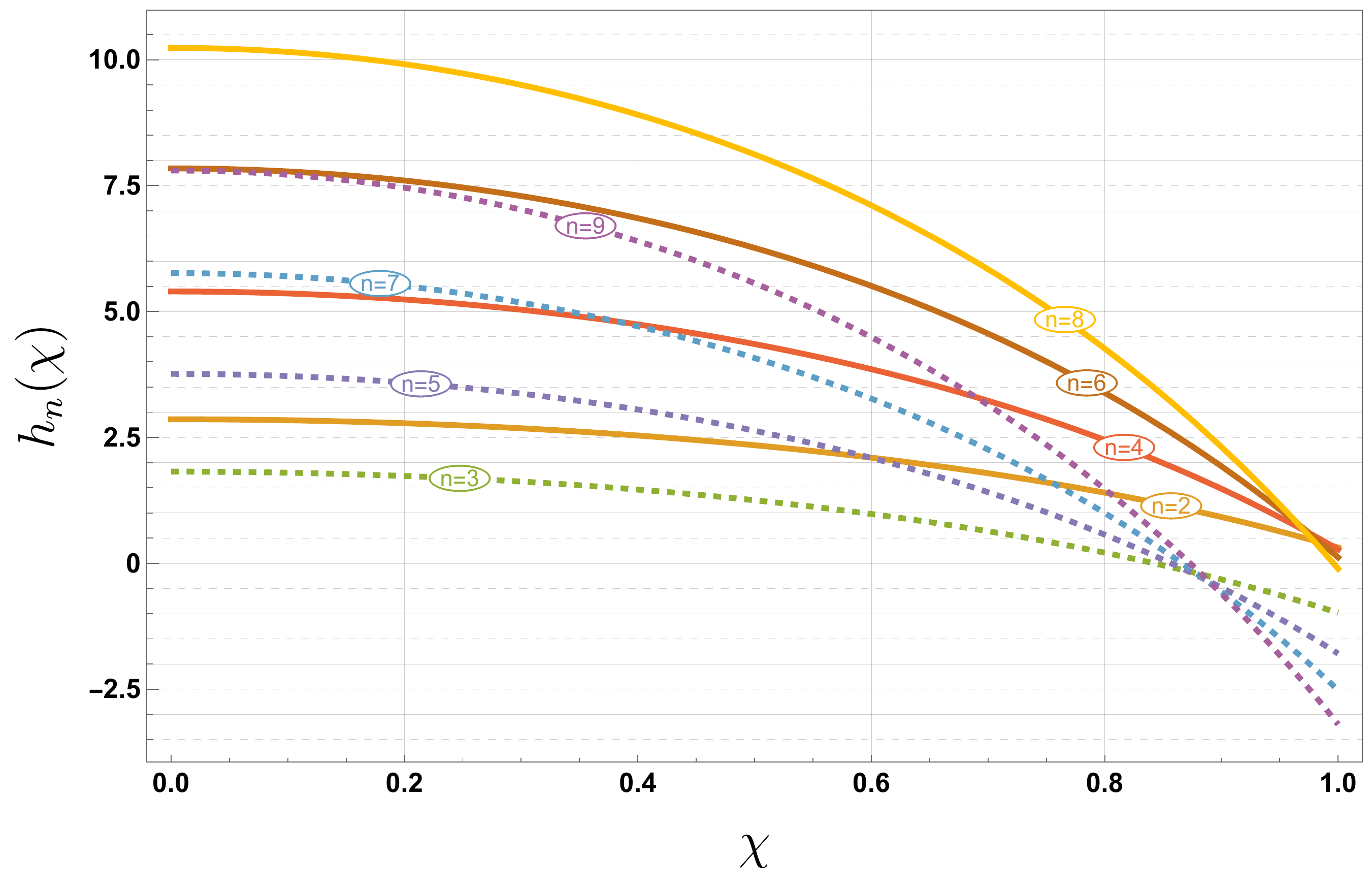}
	\caption{The relative corrections to the multipole moments $h_n$, defined in \req{eq:Znquartic}. Observe that the functions $h_n$ follow a different pattern for odd and even $n$. }
	\label{fig:hn}
\end{figure}

Using the complex function $Z_n=M_n+i S_n$, we can express the most general possible corrections as:
\begin{equation}
Z_{n}=Z_n^{(0)}\left[1+\hat\epsilon_1\,p_n(\chi)+\hat\epsilon_2\,q_n(\chi)+i\,\hat\epsilon_3\,h_n(\chi)\right]\, ,
\end{equation}
for real functions $p_n$, $q_n$ and $h_n$. However, we find that:
\begin{equation}
p_n-q_n=2h_n\, ,
\end{equation}
which seems to hold for arbitrary $n$ and angular momentum, and non-trivially is also true both for odd and even $n$. The other independent linear combination of $p_n$ and $q_n$ can be chosen to be:
\begin{equation}
p_n+q_n=:2g_n\, .
\end{equation} 
Using $g_n$ and $h_n$ we can then rewrite $Z_n$ as:
\begin{equation}\label{eq:Znquartic}
Z_{n}=Z_n^{(0)}\left[1+\left(\hat\epsilon_1+\hat\epsilon_2\right) g_n(\chi)+\left(\hat\epsilon_1-\hat\epsilon_2+i\,\hat\epsilon_3\right)h_n(\chi)\right]\, .
\end{equation}
Note that the function $g_n$ gives the corrections to the multipoles for the stringy prediction $\epsilon_1=\epsilon_2$, $\epsilon_3=0$, and hence our interest in choosing it. 
We again have constructed a series expansion in $\chi$ of these functions. In the case of $h_n(\chi)$ we are able to identify a pattern and sum the full series, finding exact results for the first few values of $n$, namely:
\begin{align}
\label{eq:h2full} h_2(\chi)&=-\frac{8}{25\,\chi^{10}}\big(64-80\chi^{10}+660\chi^8-1545\chi^6+1500\chi^4-600\chi^2\\\notag
&-8(1-\chi^2)^{5/2}(8+35\chi^4-55\chi^2)\big)\, ,\\\notag
h_3(\chi)&=\frac{8}{25\,\chi^{10}}\big(192+80\chi^{10}-620\chi^8+1165\chi^6-620\chi^4\\\notag
&-200\chi^2+8(1-\chi^2)^{5/2}(35\chi^4-35\chi^2-24)\big)\, ,\\\notag
h_4(\chi)&=\frac{8}{175\,\chi^{12}}\bigg[\sqrt{1-\chi^2}\,(3920\chi^{10}-13720\chi^8+16738\chi^6-9101\chi^4+3968\chi^2-1280)\\\notag
&+2(560\chi^{12}-4580\chi^{10}+10555\chi^8-10480\chi^6+5200\chi^4-2304\chi^2+640)\notag\\\notag
&+525\,\chi^3\arcsin(\chi)\bigg]\,,\\
h_5(\chi)&=\frac{8}{105\,\chi^{12}}\bigg[\sqrt{1-\chi^2}\,(2352\chi^{10}-7112\chi^8+5506\chi^6+1473\chi^4-7081\chi^2-3328)\notag\\
&+\big(3328+672\chi^{12}-5224\chi^{10}+9890\chi^8-5104\chi^6-2752\chi^4+12032\chi^2\big)\notag\\
&-315(5\chi^2+21)\chi\arcsin(\chi)\bigg]\, .
\label{hnexact}
\end{align}
These expressions are completely regular in the limit $\chi\rightarrow 0$, and in fact we have
\begin{align}\notag
h_2(\chi)&=
\frac{1}{400}\Big(1144-780\,\chi^2-165\,\chi^4+\mathcal{O}\big(\chi^6\big)\Big),\\\notag
h_3(\chi)&=\frac{1}{400}\Big(728-860\,\chi^2-185\,\chi^4+\mathcal{O}\big(\chi^6\big)\Big),\\\notag
h_4(\chi)&=\frac{2834}{525}-\frac{1522}{385}\chi^2-\frac{6047}{7280}\chi^4+\mathcal{O}\big(\chi^{6}\big)\, ,\\
h_5(\chi)&=\frac{1448}{385}-\frac{64109}{15015}\chi^2-\frac{19951}{21840}\chi^4+\mathcal{O}\big(\chi^{6}\big)\, .
\label{hnapprox}
\end{align}
We also observe that these series expansions converge very fast to the exact answer \req{hnexact}; Four or five terms suffice to obtain quite a precise result even for $\chi=1$. We show the first eight of these functions in Fig.~\ref{fig:hn}, where we observe that the $h_n$ for odd and even $n$ follow different patterns, although with similar behaviour, as they all decrease with $\chi$. Moreover, the plot suggests that the curves all meet at two separate points for odd and even $n$ respectively. That is not the case. We have checked it numerically up to order $\chi^{40}$, where for the multipoles on the plot and a few higher ones that is sufficient to ensure higher order corrections cannot alter the curves enough to make them all meet at the same point.

\begin{figure}[t!]
	\centering
	\includegraphics[width=\textwidth]{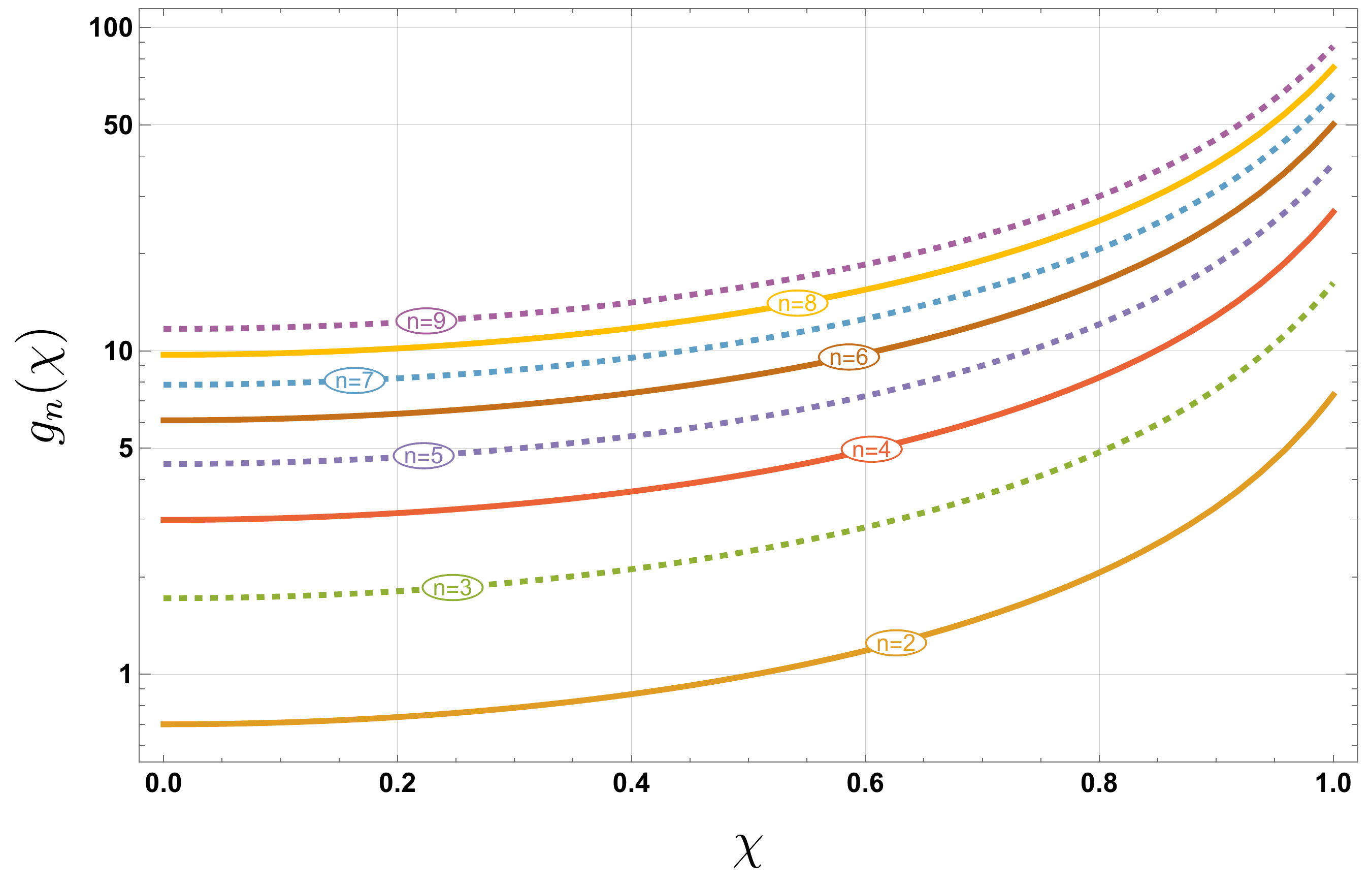}
	\caption{The relative corrections to the multipole moments $g_n$ associated to the stringy corrections $\epsilon_1=\epsilon_2$, $\epsilon_3=0$, defined in \req{eq:Znquartic}. Note that we use a log scale in the vertical axis. We observe that in this case there seems to be no distinction between odd and even $n$; the different curves seem to follow a single pattern with $n$. }
	\label{fig:gn}
\end{figure}

\begin{figure}[t!]
	\centering
	\includegraphics[width=0.5\textwidth]{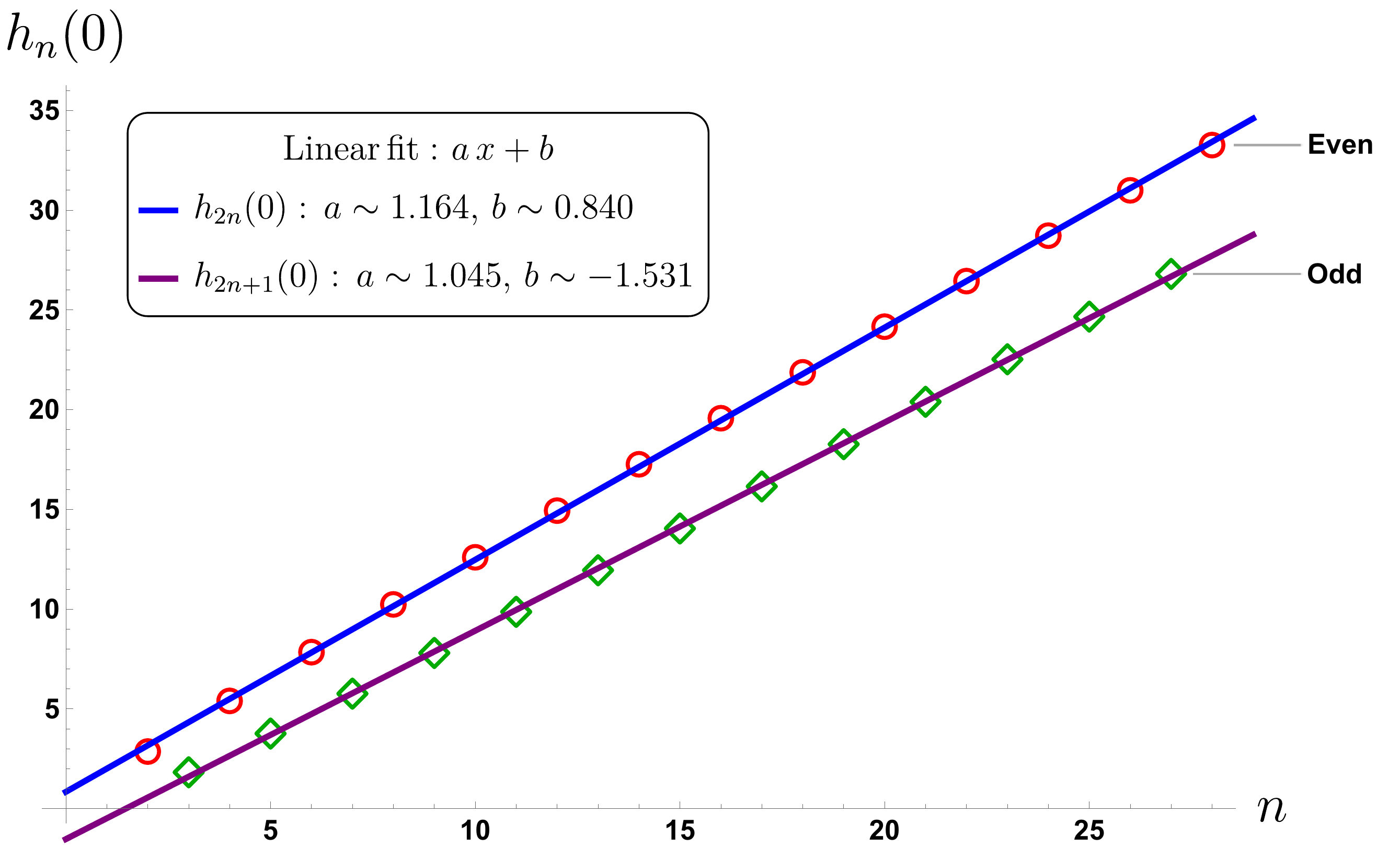}\hspace*{-1mm}
	\includegraphics[width=0.51\textwidth]{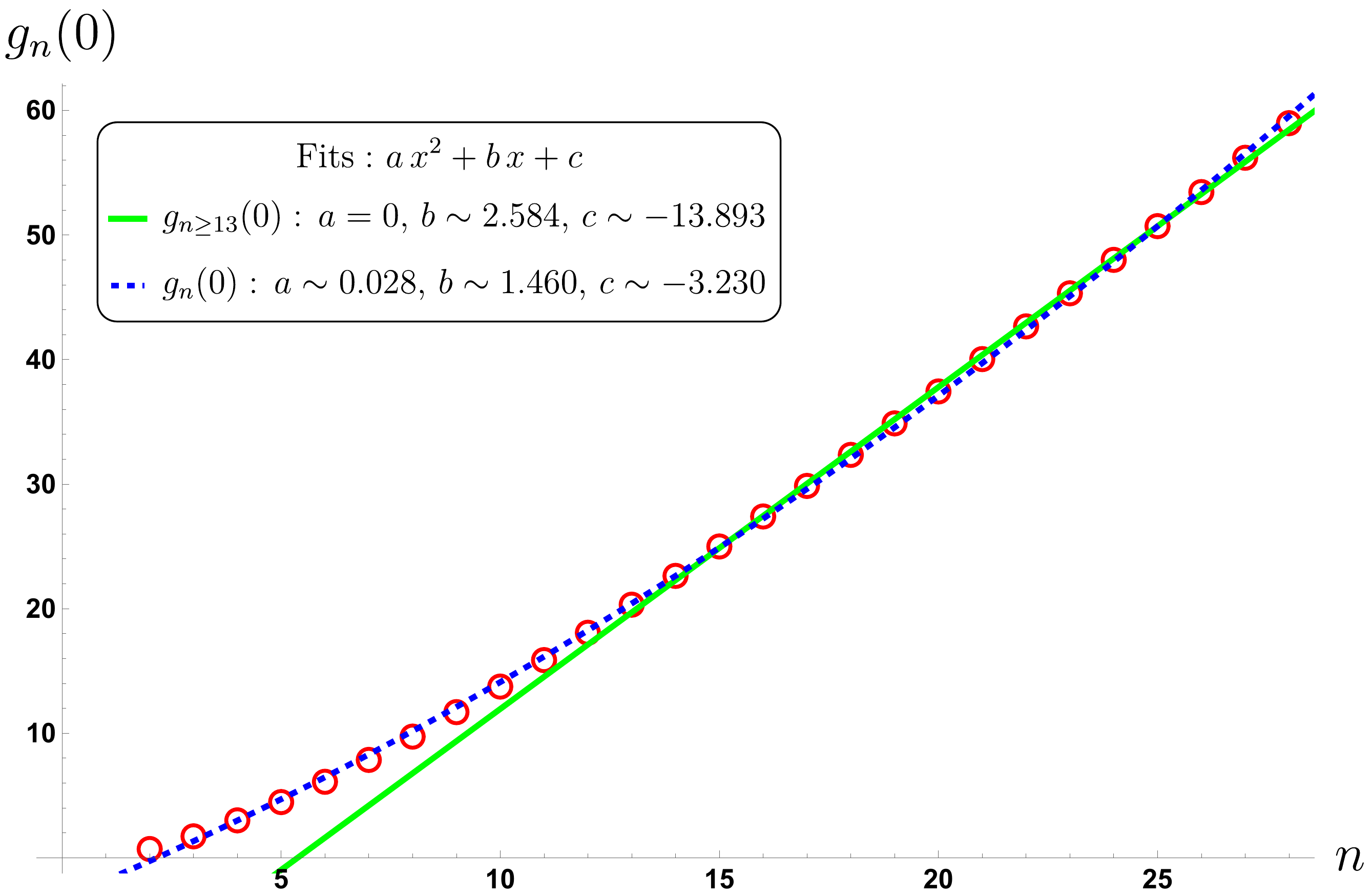}
	\caption{Fits to $h_n(0)$ and $g_n(0)$ as a function of $n$. Note that, for $g_n(0)$ the solid line is a linear fit ($a=0$) that omits the first 12 points.}
	\label{fig:quarticFits}
\end{figure}

Regarding the functions $g_n(\chi)$, we have not succeeded in finding a resummation of the power series, although we suspect this may be possible. We present here the first five terms in the series expansion, which we have computed to order\footnote{More precisely, we computed the multipole moments to order $\chi^{40}$, which means that the coefficients $g_{n}(\chi)$ are computed at order $\chi^{40-n}$.} $\chi^{40}$:
\begin{align}
\label{eq:g2pert} g_{2}(\chi)&=\frac{7}{10}+\frac{39 \chi ^2}{44}+\frac{179 \chi ^4}{208}+\frac{49 \chi ^6}{64}+\frac{818629 \chi ^8}{1244672}+\mathcal{O}(\chi^{10})\, ,\\\notag
g_{3}(\chi)&=\frac{189}{110}+\frac{109 \chi ^2}{52}+\frac{349 \chi ^4}{176}+\frac{269283 \chi ^6}{155584}+\frac{2038351 \chi
   ^8}{1391104}+\mathcal{O}(\chi^{10})\, ,\\\notag
g_{4}(\chi)&=\frac{3466}{1155}+\frac{39386 \chi ^2}{11011}+\frac{99089 \chi ^4}{29744}+\frac{6991 \chi ^6}{2431}+\frac{22064059
   \chi ^8}{9137024}+\mathcal{O}(\chi^{10})\, ,\\
g_{5}(\chi)&=\frac{1220}{273}+\frac{2244001 \chi ^2}{429429}+\frac{7288477 \chi ^4}{1516944}+\frac{51637767 \chi
   ^6}{12563408}+\frac{446380219 \chi ^8}{130202592}+\mathcal{O}(\chi^{10})\, .
\end{align}
Not only are we not able to determine the sums explicitly, but we find also that the convergence of these series is much slower than in the case of $h_n(\chi)$. Thus, we need many terms (around twenty of them for the first few $n$) in order to get an accurate answer when we approach $\chi=1$. We show the first eight $g_n$ functions in Fig.~\ref{fig:gn}. Two interesting conclusions can be drawn from this graph. First, these corrections grow fast with the spin. For example, the correction to the quadrupole, $g_2$, is around ten times larger for $\chi\sim 1$ than for $\chi\sim 0$; similar relations hold for the other $g_n$. Second, unlike the case of $h_n$, or the cubic corrections, we do not observe two different patterns for odd and even $n$ (see also Fig.~\ref{fig:quarticFits}). All the curves fit a single pattern. This seems to be a special feature of the stringy interaction \req{eq:R4rel} and it would be interesting to understand the origin of this property. 

The qualitative difference between the behaviour of $h_n$ and $g_n$ can also be inferred when considering their dependence on $n$. As in the previous section, we will only investigate the case of vanishing spin - $h_n(0)$ and $g_n(0)$. The fits are presented in Fig.~\ref{fig:quarticFits}. As for $f_n(0)$ in Fig.~\ref{fig:sixDervLinGrowth}, $h_n(0)$ exhibits a linear growth with $n$ that differs slightly for odd and even $n$. On the other hand, the growth of $g_n(0)$ does not depend on its parity. Moreover, looking at the small $n$ behaviour, one is led to perform a quadratic fit (dashed in Fig.~\ref{fig:quarticFits}). Nevertheless, the coefficient of the $x^2$ term is much smaller than that of the linear contribution and we believe that for large enough values of $n$, the growth might indeed be linear. As an example, we have also fitted a straight line (solid in Fig.~\ref{fig:quarticFits}) to the data, while omitting the first 12 points. These fits can be used to estimate the range of validity of the EFT approach for computing the multipoles, namely (using the quadratic for $\hat\epsilon_3$ case):
\begin{equation}
n_{\rm max}\sim \frac{1}{\left|\hat\epsilon_{1,2}\right|}=\frac{ M^6}{\ell^6\left|\epsilon_{1,2}\right|}\,\quad\mbox{\rm or}\quad n_{\rm max}\sim \frac{6}{\left|\hat\epsilon_3\right|}=\frac{6\,M^6}{\ell^6\left|\epsilon_3\right|},
\end{equation}
so that for $n>n_{\rm max}$ the EFT results cannot be trusted anymore. 

\subsection{Quadratic gravity}\label{sec:mult4D}
In the case of the quadratic theories \req{quadraticaction}, the metric receives parity-preserving corrections proportional to $\alpha_1^2$ and $\alpha_2^2$ and parity-breaking corrections proportional to $\alpha_1\alpha_2\sin\xi$.  Therefore, we can arrange the corrections to the complex multipoles as 

\begin{equation}\label{Znquadratic}
Z_{n}=Z_n^{(0)}\left[1+\hat{\alpha}_1^2\, a_n(\chi)+\hat{\alpha}_2^2\,b_n(\chi)+i\,\hat{\alpha}_1\,\hat{\alpha}_2\,\sin\xi\,c_n(\chi)\right]\, ,
\end{equation}
for real functions $a_n(\chi)$, $b_n(\chi)$ and $c_n(\chi)$.

\begin{figure}[t!]
	\centering
	\includegraphics[width=0.5\textwidth]{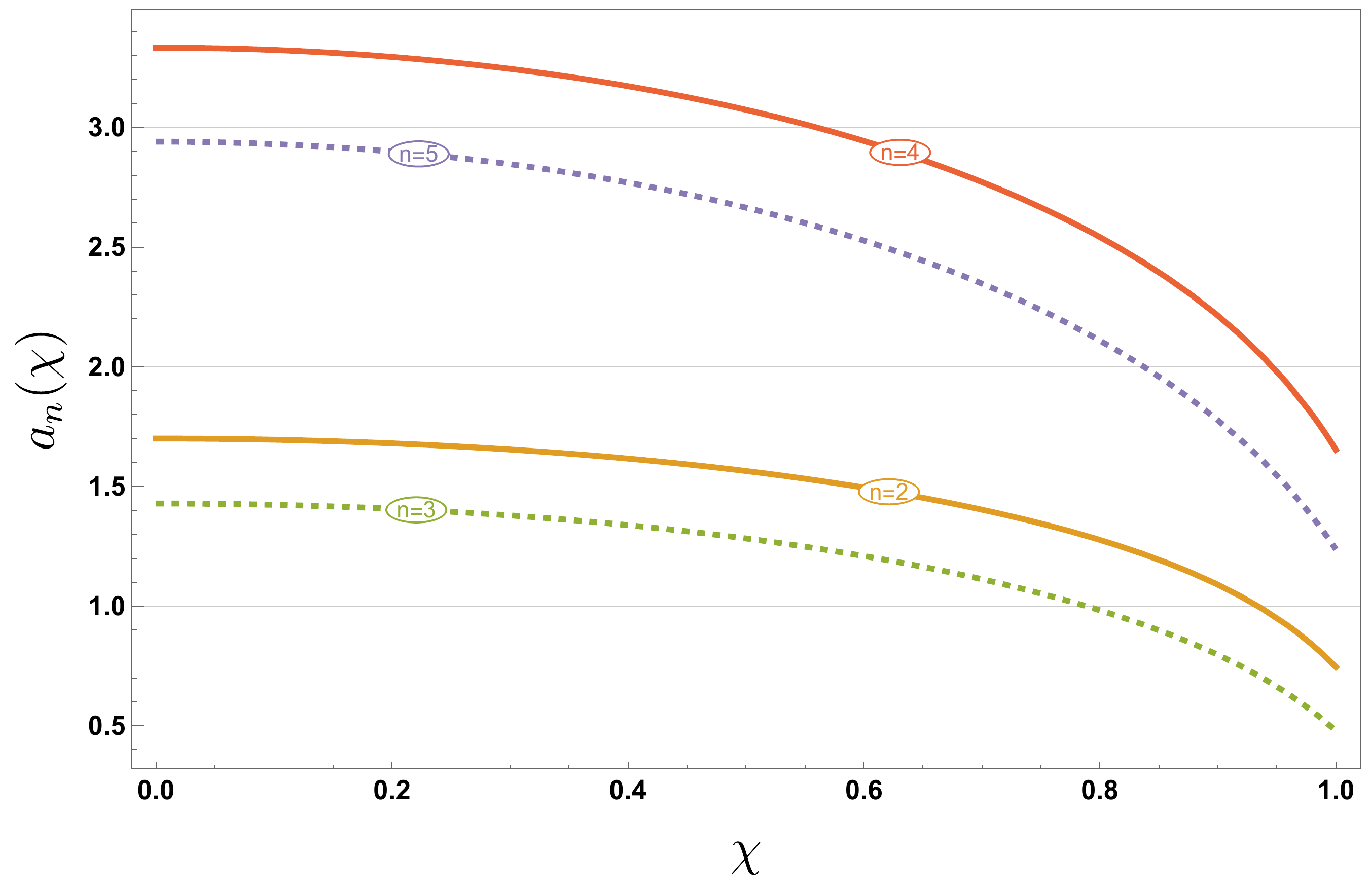}\hspace*{-1mm}
	\includegraphics[width=0.51\textwidth]{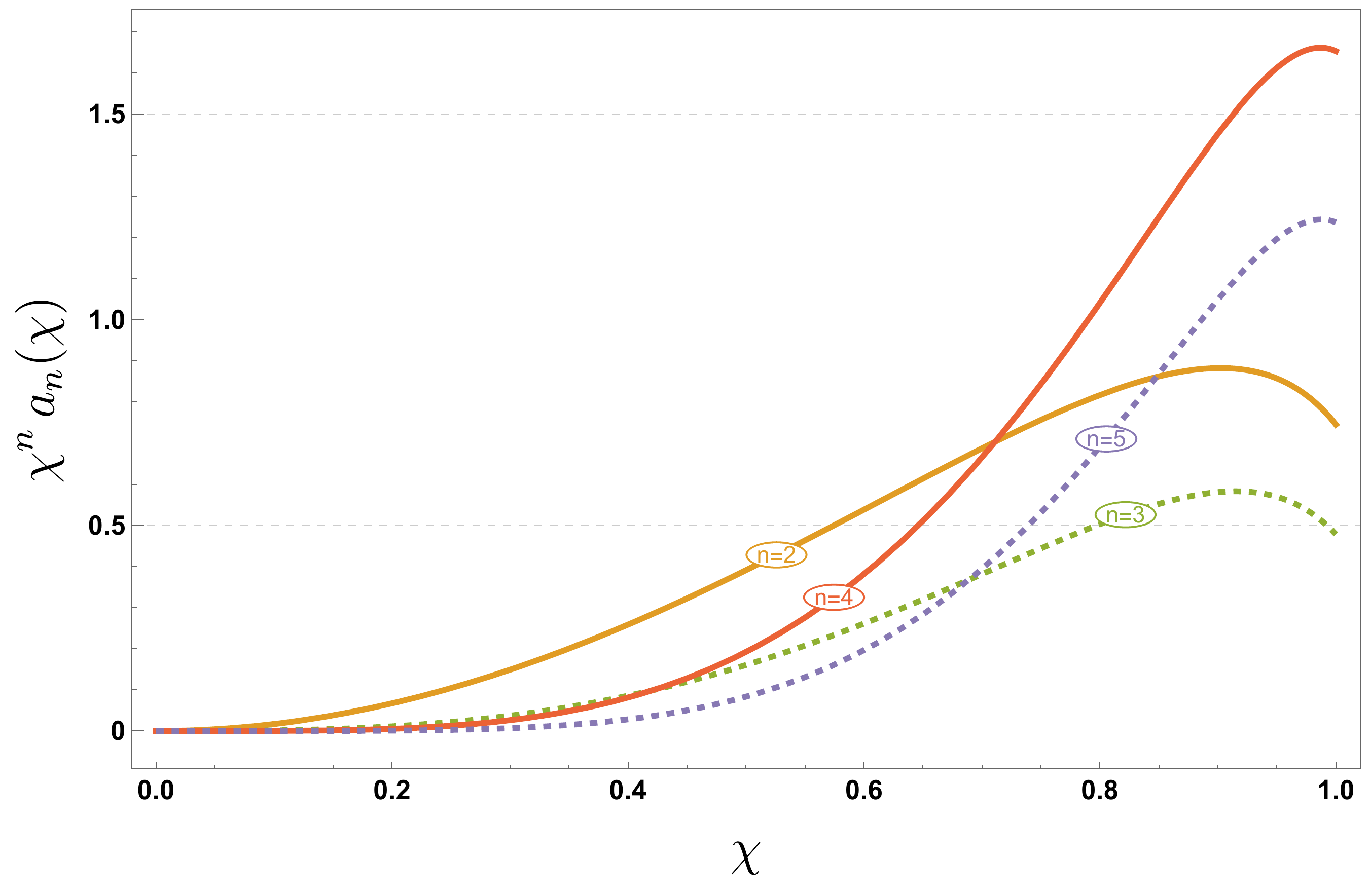}
	\includegraphics[width=0.5\textwidth]{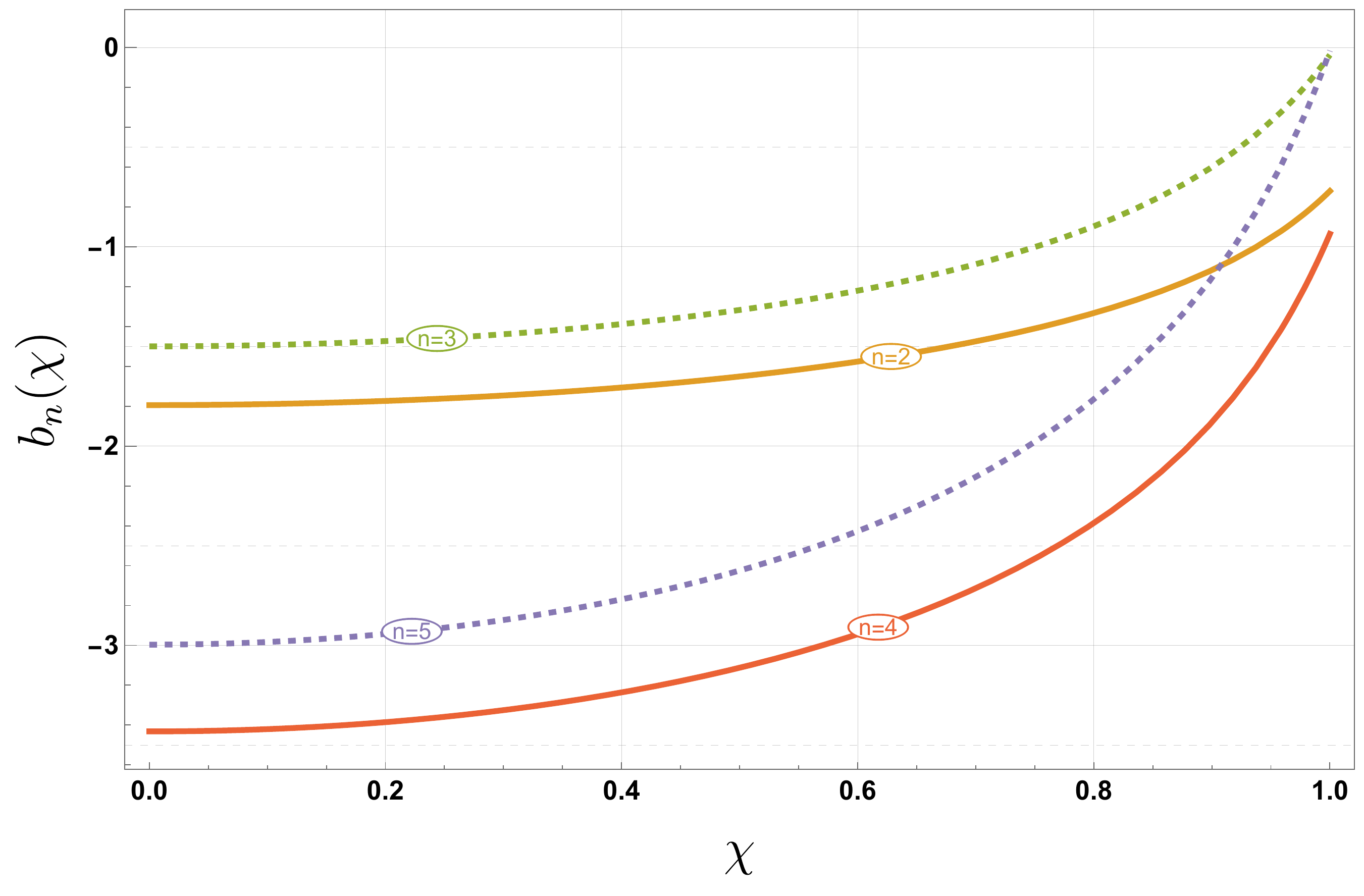}\hspace*{-1mm}
	\includegraphics[width=0.51\textwidth]{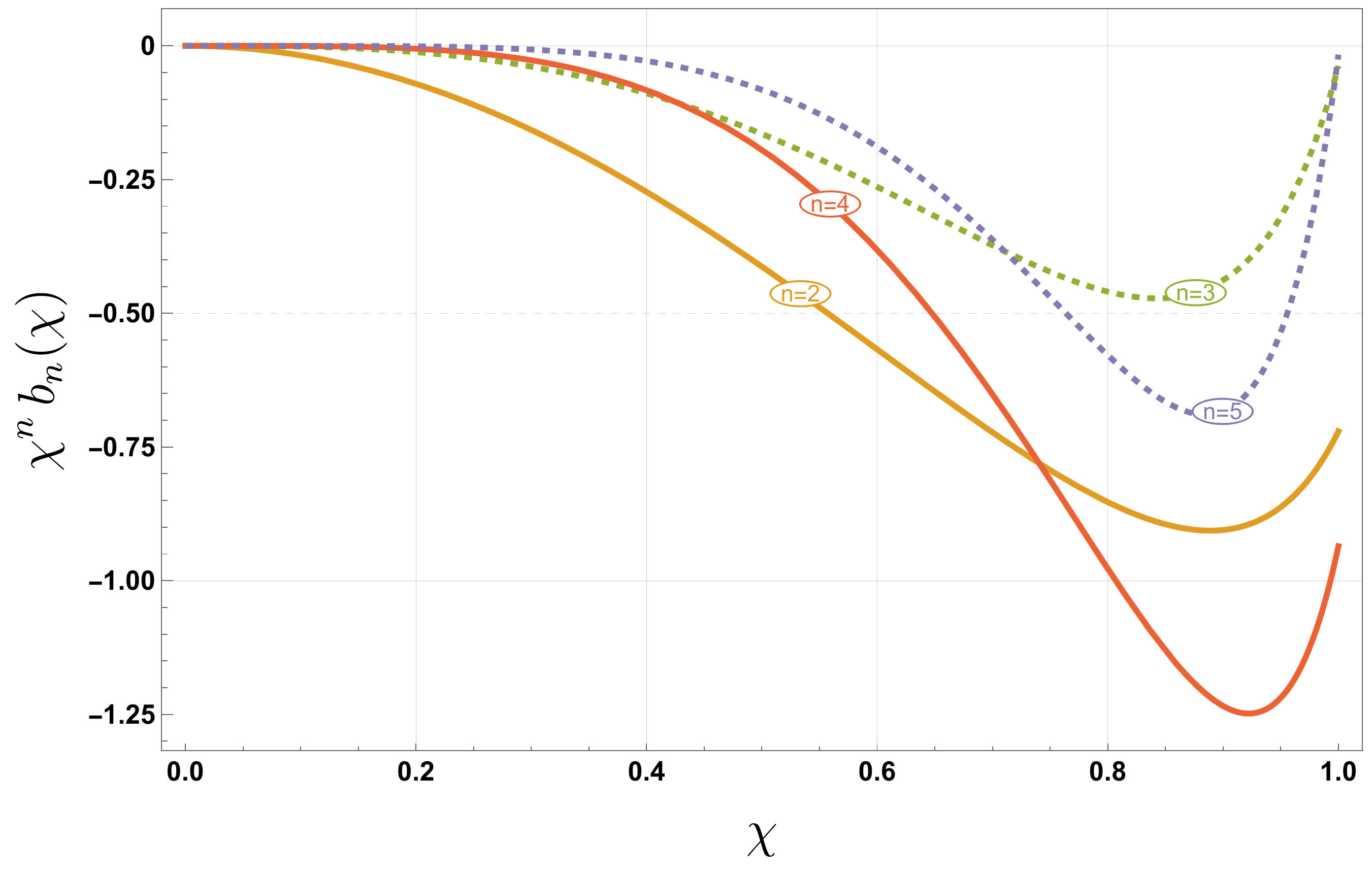}
	\includegraphics[width=0.5\textwidth]{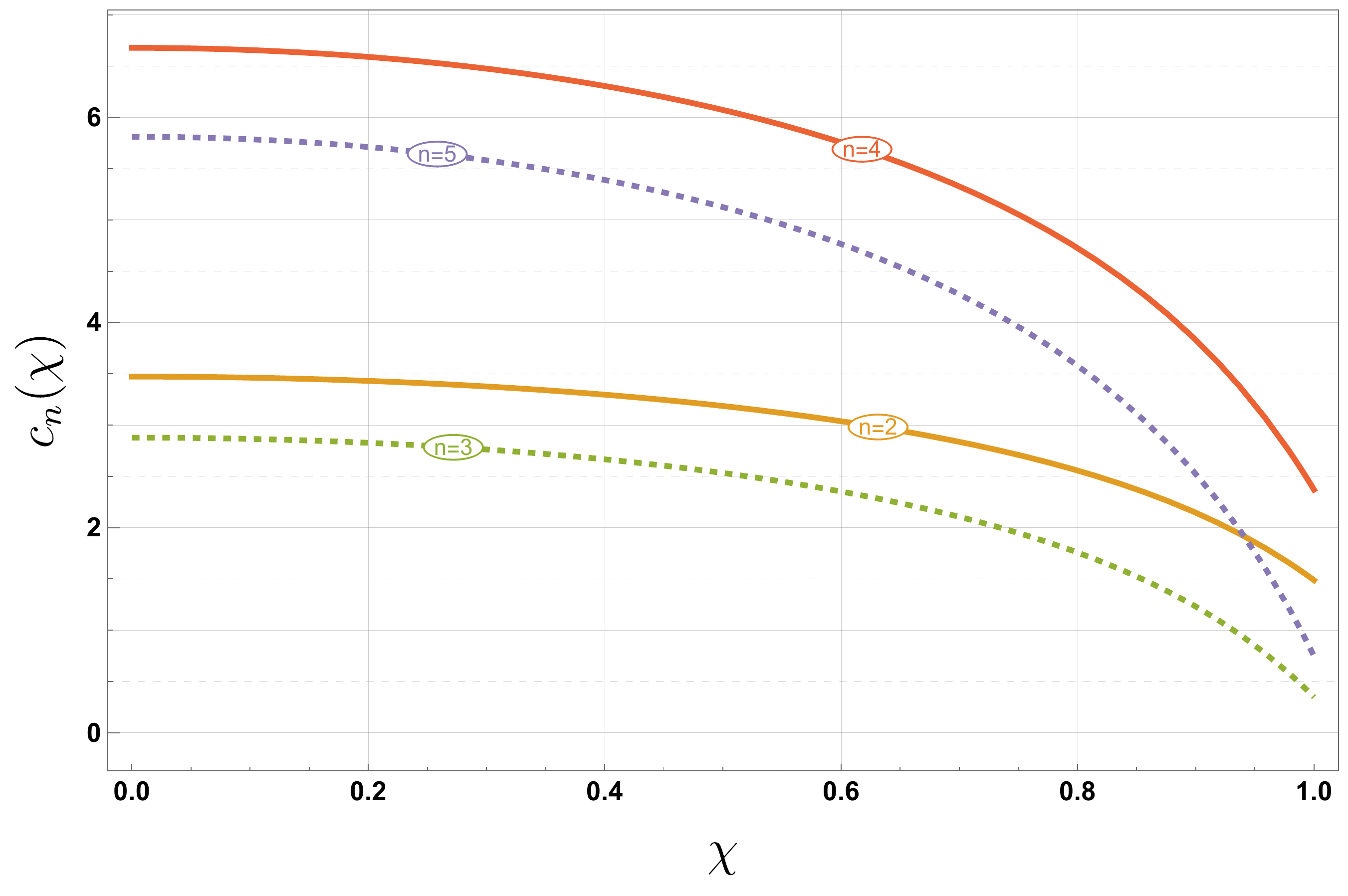}\hspace*{-1mm}
	\includegraphics[width=0.51\textwidth]{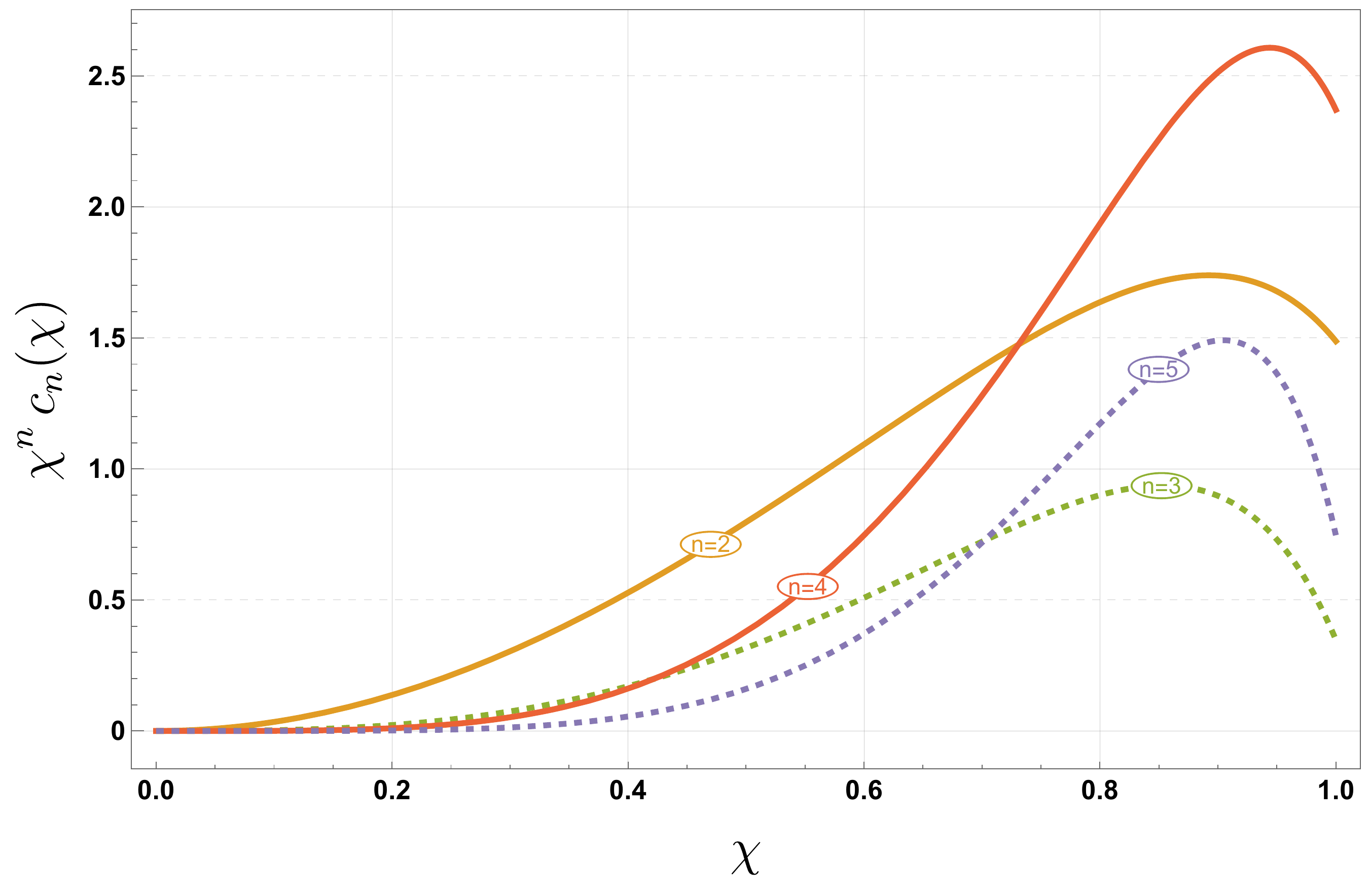}
	\caption{From top to bottom: relative (left) and absolute (right) corrections to the multipole moments associated to the dilaton-Gauss-Bonnet theory (coefficients $a_n$), dynamical Chern-Simons theory (coefficients $b_n$) and the parity-breaking interaction term  (coefficients $c_n$) . The coefficients $a_n$, $b_n$ and $c_n$ are defined in \req{Znquadratic} and represent the relative corrections to the complex multipoles $Z_n$, while the absolute correction is proportional to these coefficients times $\chi^n$. }
	\label{fig:quadratic}
\end{figure}

Unlike the case of the pure gravity EFT \req{eq:EFTofGR}, here we find no relation among these functions; all of them have different forms. In addition, we have not been able in any case to guess the exact expressions for these functions, as the coefficients of the series expansions do not seem to follow a simple pattern.
These features are unsurprising: one needs to solve the scalar equations first before being able to solve the corrections to Einstein's equations --- and no analytic solution exists even for the scalar fields.
We find the following series expansions for the first few multipole moments
\begin{align}
\label{eq:a2pert} a_2(\chi)=&\frac{4463}{2625}-\frac{33863 \chi ^2}{68600}-\frac{41760667 \chi ^4}{244490400}-\frac{9183297413 \chi
   ^6}{109880971200}+\mathcal{O}(\chi^8)\, ,\\\notag
a_3(\chi)=&\frac{26252}{18375}-\frac{991019 \chi ^2}{1852200}-\frac{43981151 \chi ^4}{256132800}-\frac{15661148969 \chi
   ^6}{204064660800}+\mathcal{O}(\chi^8)\, ,\\\notag
a_4(\chi)=&\frac{3000959}{900375}-\frac{1493088547 \chi ^2}{1568813400}-\frac{62802717559 \chi
   ^4}{208316007900}-\frac{109391232859 \chi ^6}{812432430810}+\mathcal{O}(\chi^8)\, ,\\\notag
a_5(\chi)=&\frac{314522687}{106964550}-\frac{372788157631 \chi ^2}{367102335600}-\frac{4623080093039 \chi
   ^4}{14998752568800}-\frac{286675603692757 \chi ^6}{2209816211803200}+\mathcal{O}(\chi^8)\, ,
\end{align}
\begin{align}
\label{eq:b2pert} b_2(\chi)=&-\frac{201}{112}+\frac{1819 \chi ^2}{3528}+\frac{3289259 \chi ^4}{16299360}+\frac{6651677 \chi ^6}{61471872}+\mathcal{O}(\chi^8)\, ,\\\notag
b_3(\chi)=&-\frac{8819}{5880}+\frac{3840911 \chi ^2}{5927040}+\frac{600173719 \chi ^4}{2151515520}+\frac{361612454657 \chi
   ^6}{2285524200960}+\mathcal{O}(\chi^8)\, ,\\\notag
b_4(\chi)=&-\frac{158908}{46305}+\frac{2859524347 \chi ^2}{2510101440}+\frac{603664559 \chi
   ^4}{1271350080}+\frac{661684820477 \chi ^6}{2499792094800}+\mathcal{O}(\chi^8)\, ,\\\notag
b_5(\chi)=&-\frac{146517509}{48898080}+\frac{38715214763 \chi ^2}{29368186848}+\frac{4925005246529 \chi
   ^4}{8570715753600}+\frac{17386036304479727 \chi ^6}{53035589083276800}+\mathcal{O}(\chi^8)\, ,
\end{align}
\begin{align}
c_2(\chi)=&\frac{6077}{1750}-\frac{1274269 \chi ^2}{1234800}-\frac{76837807 \chi ^4}{195592320}-\frac{6572817103 \chi
   ^6}{31394563200}+\mathcal{O}(\chi^8)\, ,\\\notag
c_3(\chi)=&\frac{20141}{7000}-\frac{9142549 \chi ^2}{7408800}-\frac{5326478623 \chi ^4}{10757577600}-\frac{780129826921 \chi
   ^6}{2856905251200}+\mathcal{O}(\chi^8)\, ,\\\notag
c_4(\chi)=&\frac{216458369}{32413500}-\frac{13654799083 \chi ^2}{6275253600}-\frac{8498104705421 \chi
   ^4}{9999168379200}-\frac{5719453119143 \chi ^6}{12379922755200}+\mathcal{O}(\chi^8)\, ,\\\notag
c_5(\chi)=&\frac{2486282089}{427858200}-\frac{450727554401 \chi ^2}{183551167800}-\frac{29620800562939 \chi
   ^4}{29997505137600}-\frac{59106789643541 \chi ^6}{108235896088320}+\mathcal{O}(\chi^8)\, .
\end{align}
These series seem to converge reasonably fast even for $\chi=1$. For instance, for $\chi=0.9$, six or seven terms seem to suffice to get an accuracy of around $1\%$, and a few more terms achieve that result for $\chi=1$.  We show these quantities as a function of the spin in Fig.~\ref{fig:quadratic}. In the case of $a_n$ and $b_n$ we used an expansion to order $\chi^{30}$, while for the parity-breaking corrections $c_n$ we have an expansion to order\footnote{This is the order of the absolute correction, this is, of $\chi^{n}a_n$, etc.} $\chi^{16}$. In all cases, the relative corrections are larger for small spin, but one has to bear in mind that the multipole moments of slowly-rotating black holes are quite small. The absolute corrections, on the other hand, have a maximum value close to extremality $\chi\sim 0.9-1$. 

We observe that the dilaton-Gauss-Bonnet corrections (coefficients $a_n$) are always positive, while those associated to dynamical Chern-Simons gravity (coefficients $b_n$) are negative. Note that our results correct the statements in \cite{Sopuerta:2009iy}, where dCS was incorrectly said to alter (only) $S_4$ and higher-order multipoles; from our results, it is clear that dCS theory alters all even-parity multipoles --- thus starting with $M_2$ --- and additionally dCS does \emph{not} break equatorial symmetry so that still $S_{2n}=M_{2n+1}=0$.

\subsection{Multipole ratios}
All of the odd-parity multipoles $S_{2n}$ and $M_{2n+1}$ vanish identically for the Kerr solution. Nevertheless, it was argued in \cite{Bena:2020see,Bena:2020uup} that certain \emph{ratios} of multipoles could be calculated for Kerr by embedding it into a larger family of string theory black holes. We briefly review this formalism introduced in \cite{Bena:2020see,Bena:2020uup} here and discuss how the ratios behave in the presence of higher-derivative corrections.\footnote{We will not discuss the more subtle ``subtracted'' ratios introduced in \cite{Bena:2020uup}, which also involve vanishing ratios involving (only) the even parity multipoles in the numerator.}

In \cite{Bena:2020see,Bena:2020uup}, the Kerr black hole is embedded in the most general family of non-extremal, rotating black holes in four-dimensional STU supergravity \cite{Chow:2014cca}. This black hole has ten parameters --- four electric and four magnetic charges in addition to the mass and angular momentum. A generic black hole of this kind has all its multipoles non-vanishing. These multipoles themselves can be seen as functions of four variables $M,J,D,a$ --- resp. the mass $M$, angular momentum $J$, the so-called dipole parameter $D$, and rotation parameter $a$; note that in general $M$, $D$, and $J$ depend in a complicated way on the eight electromagnetic charges so that in particular $J\neq M a$. We can now construct \emph{ratios} of multipoles $\mathcal{M}(M,J,D,a)$ and take the Kerr limit ($D\rightarrow 0$ and $J\rightarrow M a$); if this limit is well-defined, this limit defines a new \emph{multipole ratio for Kerr}, even if the multipoles involved in the ratio vanish on the Kerr solution.

As an example, consider:
\be \mathcal{M} = \frac{M_{\ell+1}M_{\ell+2}}{M_\ell M_{\ell+3}}.\ee
On the Kerr solution, for every $\ell$ the numerator \emph{and} denominator vanish, so this ratio is ill-defined. However, the above procedure gives a well-defined limit, and one finds \cite{Bena:2020uup}:
\be \label{eq:ratio1} \mathcal{M}^\text{(Kerr)} = \lim_{D\rightarrow 0, J\rightarrow Ma} \mathcal{M}(M,J,D,a) = 1 - \frac{4}{3+(-1)^\ell (2\ell+1)}.\ee
Another example is:
\be \label{eq:ratio2} \frac{M_2 S_\ell}{M_{\ell+1} S_1} = 1,\ee
which is trivially true for Kerr when $\ell$ is odd, but requires the above limiting procedure for even $\ell$. Further examples of multipole ratios for Kerr can be found in \cite{Bena:2020see,Bena:2020uup}.

\subsubsection*{Multipole ratios with higher-derivative corrections from string theory}

In \cite{Bena:2020see,Bena:2020uup}, it was conjectured that these Kerr multipole ratios such as (\ref{eq:ratio1}) were a string theory prediction; small deviations from Kerr are constrained by the demand that these ratios remain (well-defined and) the same value. More precisely, these ratios should be a prediction of string theory compactified on a torus to four dimensions.

In such a toroidal compactification, we discussed above that the string-theory higher-derivative corrections are given by (\ref{eq:STcorr}). In particular, there are no odd-parity corrections, so that the odd-parity multipoles $S_{2n}$ and $M_{2n+1}$ remain zero. This gives a perhaps rather unsatisfying ``confirmation'' of the conjecture in \cite{Bena:2020see,Bena:2020uup} --- the perturbations to the Kerr solution due to string theory higher-derivative corrections leave the multipole ratios invariant since the odd-parity multipoles remain vanishing.\footnote{Of course, a much more powerful check would be to calculate the (string-theoretic) higher-derivative corrections to the most general STU black hole (i.e. with all multipoles non-vanishing) and then re-calculate the multipole ratios using these corrected black holes. Calculating the higher-derivative corrections to this general black hole would be an interesting challenge.}

\subsubsection*{Multipole ratios with general higher-derivative corrections}
It is relatively easy to see that the multipole ratios of \cite{Bena:2020see,Bena:2020uup} will \emph{not} remain unaltered when generic higher-derivative corrections are turned on --- in particular, when odd-parity higher-derivative corrections are present. For example, when the odd-parity six-derivative parameter $\lambda_{\rm odd}\neq0$, the ratio (\ref{eq:ratio2}) becomes, for $\ell=2n$:
\be \label{eq:ratio2new} \left(\frac{M_2 S_{2n}}{M_{2n+1} S_1}\right)_{\lambda_{\rm odd}\neq0} = \frac{f_{2n}(\chi)}{f_{2n+1}(\chi)}.\ee
From (\ref{eq:6Df2full})-(\ref{eq:6Df4full}) or (\ref{eq:6Df2pert})-(\ref{eq:6Df4pert}), it is clear that this ratio is no longer equal to 1 as in (\ref{eq:ratio2}). Note that the ratio (\ref{eq:ratio2new}) is independent of $\lambda_{\rm odd}$ even though its calculation requires $\lambda_{\rm odd}\neq 0$.

Similarly, for (\ref{eq:ratio1}) with $\ell=2n$, we now find:
\be  \label{eq:ratio1new} \left(\frac{M_{2n+1}M_{2n+2}}{M_{2n} M_{2n+3}}\right)_{\lambda_{\rm odd}\neq0} = \frac{f_{2n+1}(\chi)}{f_{2n+3}(\chi)},\ee
which again does not match with the Kerr value (\ref{eq:ratio1}). A similar analysis could be made for other multipole ratios, as well as for the eight-derivative odd-parity corrections when $\epsilon_3\neq 0$ or the four-derivative odd-parity corrections when $\alpha_1\alpha_2\sin\xi\neq 0$.

\section{Observability}
\label{sec:obs}

As mentioned in the Introduction in Section \ref{sec:Introduction}, the observation of gravitational waves coming from binary black hole merger events provides an exciting new opportunity to measure and constrain gravity effects beyond general relativity. This includes constraining the scale of possible higher-derivative corrections to new levels of precision, mostly through the perturbations that these corrections imply to the structure of the merging black holes --- which is encoded in the multipole moments we have calculated in Section \ref{sec:BHmultipoles}.

Of course, the constraints we are able to put on the higher-derivative length scale $\ell$ will still be (many) orders of magnitude away from the Planck scale, $\ell_\text{Pl} \sim 10^{-35}\, \text{m}$, which is a priori the natural scale at which to expect such higher-derivative corrections. However, it is also clear that extensions of general relativity at scales much larger than the Planck scale are not ruled out by current experiments and observations; it is important to understand the extent to which current and future gravitational wave experiments will further constrain the available phase space of effective field theories beyond GR \cite{Cardoso:2018ptl}, by adopting a theory-agnostic viewpoint and without a priori limiting ourselves with a theoretical bias of naturalness \cite{Sennett:2019bpc}.

A binary black hole merger consists of three phases. First, there is the relatively long \emph{inspiral} phase, where the black holes are in orbit around each other. This phase transitions into the violent and short \emph{merger} phase, where the black hole horizons coalesce into a single object. Finally, the new object relaxes to a (quasi-)stationary state in the \emph{ringdown} phase.

We will focus on the inspiral phase --- when the black holes are still sufficiently far away from each other that typically a post-Newtonian expansion is possible to describe the orbit evolution. Of course, the strong gravity merger phase will most likely be even more sensitive to higher-derivative corrections, but this is unfortunately hard to calculate. Higher-derivative corrections to the ringdown phase --- the relaxation to stationarity of the final black hole --- were studied in \cite{Cardoso:2018ptl,Cano:2020cao,Pierini:2021jxd,Wagle:2021tam,Srivastava:2021imr,Cano:2021myl}. We will assume the higher-derivative scale $\ell$ is small compared to the black hole scale(s) --- so $\ell\leq M$ for any black hole mass $M$ involved.

\subsubsection*{Leading higher-derivative corrections to inspiral dynamics}

The presence of higher-derivative  terms corrects the gravitational dynamics in two ways:
\begin{itemize}
 \item[(A)] finite-size effects --- i.e. each inspiralling black hole (individually) has corrections to its multipole structure;
 \item[(B)] the coupling of the inspiralling system to the gravitational field is changed, correcting the resulting gravitational wave radiation.
\end{itemize}
The corrections in the coupling to radiation of (B) can be estimated from the energy dissipation:
\be \frac{dE}{dt} =-\frac15 \langle \dddot{Q}_{ij}\dddot{Q}_{ij}\rangle,\ee
with the effective quadrupole coupling with non-zero eight-derivative terms given by \cite{Endlich:2017tqa}:
\be \label{eq:Qijdiff} Q_{ij} = Q_{ij}^{(0)}\left(1 + c_{rad} \frac{\epsilon_1\ell^6}{r^6} \left(\frac{M}{r}\right)^2\right),\ee
where $M=M_1+M_2$ is the total mass of the binary system, $Q_{ij}^{(0)}$ is the (two-derivative) regular mass quadrupole of the binary system, and
$c_{rad}$ is a numerical factor. 
(Note that $\epsilon_{2,3}$ do not contribute to the shift of $Q_{ij}$, although they do shift the current quadrupole of the system \cite{Endlich:2017tqa}.)
In the inspiral, we can expand the system's evolution in powers of the dimensionless angular velocity of the orbit $v$ (with $v = (2\pi M \nu)^{1/3}$ where $\nu$ is the frequency of the orbit).
Then $r\sim M/v^2$, so that the corrections to this effective quadrupole $Q_{ij}$ scale as $\delta Q_{ij} \sim Q_{ij}^{(0)}(\epsilon_1\ell^6/M^6) v^{16}$ and so $\delta(dE/dt)\sim \mathcal{O}(v^{26})$. Presumably (although not discussed in \cite{Endlich:2017tqa}), the analogous six-derivative corrections would then scale as $\delta(dE/dt)\sim \mathcal{O}(v^{22})$.

By contrast, for the finite-size effects of (A), the leading order effect is due to the change of the mass quadrupole $M_2$, which enters at $\mathcal{O}(v^{14})$ \cite{Ryan:1995wh,Poisson:1997ha} in $dE/dt$ for the inspiral.
(Note that a non-zero $S_2$ enters at $\mathcal{O}(v^{15})$ in $dE/dt$ \cite{Fransen:2022jtw}.) So, we expect the finite-size effects of (A) to be dominant over the corrections to the radiation coupling of (B). We have derived the modifications to the mass quadrupole due to the various possible higher-derivative terms in Section \ref{sec:BHmultipoles}. These can be used to estimate the observability of the higher-derivative length scale $\ell$, which will be carried out further down.

Finally, note that our analysis above is for eight- and six-derivative corrections. The situation is different for four-derivative terms: for example, note that from (\ref{Znquadratic}), the changes to the black hole multipoles are \emph{quadratic} in the effective coupling $\alpha_i \ell^2/M^2$, which means they behave more like a six-derivative correction to the multipole structure. Certain observable aspects of particular four-derivative additions were discussed in the context of dynamical Chern-Simons theory in e.g. \cite{Sopuerta:2009iy,Yagi:2012vf}, and of Einstein-dilaton-Gauss-Bonnet in \cite{Wang:2021yll}. Both dCS and EdGB were analyzed in \cite{Perkins:2021mhb}.
We will only consider the eight- and six-derivative corrections in the rest of this section, except when comparing to the existing bounds on the four-derivative length scale at the end.

\subsubsection*{Constraining the higher-derivative scale $\ell$}
The possible deviations from the Kerr mass quadrupole $M_2$, for each of the two black holes in a binary system, are often parametrized as:
\be M_{2,(i)} = -\kappa_{(i)} M_{(i)}^3 \chi_{(i)}^2, \qquad \kappa_{(i)} = 1 + \delta\kappa_{(i)},\ee
where $\delta\kappa_{(i)} = 0$ is the GR prediction. 
One typically forms the symmetric and antisymmetric combinations $\delta\kappa^{(s)}=(1/2)(\delta\kappa_{(1)}+\delta\kappa_{(2)})$ and $\delta\kappa^{(a)}=(1/2)(\delta\kappa_{(1)}-\delta\kappa_{(2)})$, as these are better suited for measurements.

We can calculate the expectation for $\delta \kappa^{(s)}$ from our higher-derivative corrections.
From (\ref{eq:zn6Derv}), (\ref{eq:Znquartic}), and (\ref{Znquadratic}) above, we have:
\be\label{eq:kappainHD} \delta \kappa^{(s)} = \left( \left[ \tilde\alpha_1^2\, a_2(0)+\tilde\alpha_2^2\,b_2(0) \right] + \tilde\lambda_\text{ev}\, f_2(0) + \left[(\tilde\epsilon_1+\tilde\epsilon_2)\,g_2(0) + (\tilde\epsilon_1-\tilde\epsilon_2)\,h_2(0)\right]\right) + \mathcal{O}\left(\chi_{(1)}^2,\chi_{(2)}^2\right),\ee
where $f_2(\chi)$ is given in (\ref{eq:6Df2full}), $g_2(\chi), h_2(\chi)$ in (\ref{eq:g2pert}) and (\ref{eq:h2full}), and $a_2(\chi)$, $b_2(\chi)$ in (\ref{eq:a2pert}) and (\ref{eq:b2pert}); note that $a_2(0),b_2(0),f_2(0),g_2(0),h_2(0)$ are all $\mathcal{O}(1)$ numbers. 
Finally, the tilded quantities should be understood in (\ref{eq:kappainHD}) as an appropriate mean over inverse powers of the masses, so e.g.:
\be \tilde\lambda_\text{ev} = \frac12\left(\frac{1}{M_1^4} + \frac{1}{M_2^4}\right)\ell^4\lambda_\text{ev},\ee
which can be compared to the definitions of the effective couplings for a single black hole given in (\ref{eq:dimlessallcouplings}).

The asymmetric combination $\delta\kappa^{(a)}$ can be compared to (\ref{eq:kappainHD}) as:
\be \label{eq:deltakappaanti} \delta \kappa^{(a)} =  2\frac{\delta M}{M_1}\delta \kappa^{(s)}  + \mathcal{O}\left(\chi_{(1)}^2,\chi_{(2)}^2,(\delta M)^2\right),\ee
where $\delta M = M_2-M_1$. For (approximately) equal mass binaries and low spins, then, $\delta \kappa^{(a)}$ will be at most the same order as $\delta \kappa^{(s)}$ for our higher-derivative corrections.

In practice, one often constrains the symmetric combination $\delta\kappa^{(s)}$, assuming the antisymmetric one vanishes, $\delta\kappa^{(a)} = 0$ \cite{Krishnendu:2017shb,LIGOScientific:2021sio}.
 It is also possible to leave both combinations as free parameters; in this case, the constraints on $\delta\kappa^{(a)}$ are much weaker than on $\delta\kappa^{(s)}$ \cite{Krishnendu:2018nqa}.
We will focus on $\delta\kappa^{(s)}$ as a good estimate of the measurability of the higher-derivative corrections to the quadrupole $M_2$. From (\ref{eq:kappainHD}) we can conclude that, at least for relatively low spin, the (order of magnitude) constraint on $\delta\kappa^{(s)}$ translates into a spin-independent constrain on $\ell$ once the binary black hole masses $M_{(i)}$ are known.
 However, we do note that the often-used assumption that $\delta\kappa^{(a)}=0$ is clearly not optimal when considering higher-derivative corrections. In fact, it would improve the measurability and constrainability of the higher-derivative corrections to repeat the analysis constraining the multipole deviations, by considering both non-zero $\delta\kappa^{(s)}$ \emph{and} $\delta\kappa^{(a)}$, but where moreover the values of both these corrections are linked through an equation such as (\ref{eq:deltakappaanti}).

The best bound on $\delta\kappa^{(s)}$ with current observations is roughly $-16.0\lesssim \delta\kappa^{(s)}\lesssim 6.66$ \cite{LIGOScientific:2021sio},\footnote{The constraint on negative values of $\delta\kappa^{(s)}$ is worse than the bound on positive values; this is due to how the parameters correlate with the effective binary spin parameter \cite{LIGOScientific:2021sio}.} which gives:
\be \label{eq:preconstraint} -18.67 \left[ \frac12\left(\frac{1}{M_1^4} + \frac{1}{M_2^4}\right)\right]^{-1} \lesssim \lambda_\text{ev}\, \ell^4 \lesssim 7.77 \left[ \frac12\left(\frac{1}{M_1^4} + \frac{1}{M_2^4}\right)\right]^{-1},\ee
where we took only $\lambda_\text{ev}\neq 0$ in (\ref{eq:kappainHD}) for simplicity.
This constraint is for binary black hole masses of $M=1-10\,M_\odot$.\footnote{Note that the solar mass is approximately $M_\odot\approx 1.5\, \text{km}$.} Assuming also the dimensionless coupling in the Lagrangian $\lambda_\text{ev}$ is a positive $\mathcal{O}(1)$ number, we get (in a best case rough estimate):
\be \label{eq:weakconstraintell} \ell \lesssim 1.67\left[ \frac{\lambda_\text{ev}}{2}\left(\frac{1}{M_1^4} + \frac{1}{M_2^4}\right)\right]^{-1/4} \sim 1-10\, \text{km}.    \ee
Note that $\lambda_\text{ev}<0$ would lead to a less stringent constraint through the negative constraint in (\ref{eq:preconstraint}).
Considering other non-zero higher-derivative corrections gives comparable constraints on $\ell$ from (\ref{eq:kappainHD}).

At the Einstein Telescope, a future third-generation ground-based detector \cite{Krishnendu:2018nqa}, the bound is estimated to improve roughly two orders of magnitude for similar-sized black hole mergers,  $\delta\kappa^{(s)}\lesssim 10^{-2}$. Again considering only $\lambda_\text{ev}\neq 0$ (and $\lambda_\text{ev}>0$), this would translate to a bound on $\ell$ of:
\be\label{eq:bestconstraintell} \ell \lesssim 0.32 \left[ \frac{\lambda_\text{ev}}{2}\left(\frac{1}{M_1^4} + \frac{1}{M_2^4}\right)\right]^{-1/4} \sim 0.1-1\, \text{km} .\ee
This is the best constraint possible in the near future.

\subsubsection*{Comparison with other constraints}
Here, we briefly list a few other current or future observational aspects of gravitational waves which can be used to constrain $\ell$.

First of all, note that \cite{Perkins:2021mhb} provides the best constraints to date for the length scale of the four-derivative Einstein-dilaton-Gauss-Bonnet and dynamical Chern-Simons theories: $\ell\lesssim 1.7\, \text{km}$ for EdGB, and $\ell\lesssim 8.5 \,\text{km}$ for dCS (see also \cite{Silva:2020acr}). Note that these are already comparable to our six- or eight-derivative constraints (\ref{eq:bestconstraintell}) for the third-generation detectors, even though the four-derivative scale constraints in \cite{Perkins:2021mhb} come from \emph{current} detections. It is reasonable to assume that third-generation detectors will be able to constrain four-derivative theories much better than (\ref{eq:bestconstraintell}).

Perhaps most interesting to contrast with the equal- and relatively low-mass binary black hole mergers discussed above, are the extreme-mass ratio inspirals of small solar-mass objects into supermassive black holes (of masses $\sim 10^5-10^7\, M_\odot$), that will be detected at the future space-based detector LISA. These will be able to constrain deviations from the Kerr prediction for the \emph{dimensionless} mass quadrupole  $M_2/M^3$ of the supermassive partner extremely well \cite{Barack:2006pq, Gair:2017ynp}: up to $\Delta (M_2/M^3)\sim 10^{-4}$.\footnote{Note that this is roughly of the same order of magnitude as the ET bound $\delta\kappa^{(s)}\sim 10^{-2}$ if $\chi\sim 10^{-1}$.}
However, since $M$ here is the mass of a supermassive black hole, $M\sim 10^5 M_\odot$ or higher, this translates to a rather poor constraint on $\ell$ itself: 
\be \ell \lesssim 0.1 (\lambda_\text{ev}\chi^2)^{-1/4} M \sim 10^4 \, \text{km},\ee
where we assume the best case of a highly-spinning supermassive black hole, so that $\chi \sim 1$.
The dimensionless current quadrupole can similarly be constrained by EMRIs at LISA to $\Delta (S_2/M^3)\sim 10^{-2}$ \cite{Fransen:2022jtw}; this would lead to slightly worse constraints on $\ell$.

Black holes in a binary system are tidally deformed by each others' gravitational field --- this deformation is characterized by tidal Love numbers, which for the Kerr black hole (in two-derivative Einstein gravity) identically vanish \cite{Charalambous:2021mea}. In the presence of eight-derivative corrections, it was calculated in \cite{Cardoso:2018ptl} that the quadrupolar tidal Love numbers become $k_2^{E,B} \sim \mathcal{O}(10)\tilde\epsilon_{1,2}$. In \cite{Cardoso:2017cfl}, current detectors are estimated to constrain $k_2^E\lesssim 100$; LISA could do better at $k_2^E\lesssim 10$,\footnote{In \cite{Pani:2019cyc}, it was estimated that EMRIs at LISA may actually be able to constrain $k_2^E\lesssim 10^{-5}$; however, the corresponding bound on $\ell$ would be again suppressed by the mass of the supermassive black hole.} and (optimistically) ET could constrain an extra two orders of magnitude compared to aLIGO/aVIRGO, down to $k_2^E\lesssim 1$. In this latter case, we would have the constraint:
\be \ell \lesssim \epsilon_i^{-1/6} M\sim 1\, \text{km},\ee
which is comparable to (\ref{eq:bestconstraintell}).

Constraining the higher-derivative length scales through the measurement of the perturbed quasinormal modes in the ringdown phase was considered in \cite{Cardoso:2018ptl} for eight-derivative corrections. They estimate that roughly $\ell\lesssim 10\, \text{km}$ is the best that aLIGO/aVIRGO measurements will be able to do for such ringdown measurements. The more detailed analysis of \cite{Cano:2020cao} suggests a constraint of e.g. $|\hat\lambda_\text{ev}|\leq 0.1$ at ET from quasinormal mode observations; this is again comparable to (\ref{eq:bestconstraintell}).

In our analysis, we have assumed that the higher-derivative scale is \emph{small}, and in any case smaller than the corresponding black hole scale(s), so $\ell\lesssim M_{(i)}$; note that the weak bounds such as (\ref{eq:weakconstraintell}) are (at best) at the boundary of this regime. For $\ell \gtrsim M_{(i)}$, one instead expects the finite-size effects to become subleading to the corrections to the gravitational wave radiation \cite{Sennett:2019bpc}. Eight-derivative corrections in this regime were considered in \cite{Sennett:2019bpc}, where the conclusion was that roughly the range $100\, \text{km} \lesssim \ell \lesssim 200\, \text{km}$ is strongly disfavored.

Finally, we wish to mention the analysis of \cite{Kastha:2018bcr,Kastha:2019brk}. They parametrize deviations to the \emph{radiated} multipole moments of the binary system; for example, $Q_{ij} = \mu_2 Q_{ij}^{(0)}$, where $\mu_2=1$ in GR. They analyze the constraints on e.g. $\delta\mu_2=\mu_2-1$. Relating $\delta\mu_2$ to the higher-derivative corrections requires knowledge of the corrected black hole multipoles as well as correction effects to the gravitational wave propagating from the source to the detector onto the curved spacetime. Systematically characterising the contributions to $\delta\mu_2$ coming from higher-derivative terms in the action might offer interesting additional constraints on the length scales of theories beyond GR.

\section*{Acknowledgments}
We thank I. Bena, N. Bobev, G. Comp\`ere, K. Fransen, T. Hertog and D. Pereñiguez for interesting discussions.
The work of PAC is supported by a postdoctoral fellowship from the Research Foundation - Flanders (FWO grant 12ZH121N).
BG is supported in part by the ERC Grant 787320 - QBH Structure.
DRM is supported by FWO Research Project G.0926.17N.
AR is supported by a postdoctoral fellowship associated
to the MIUR-PRIN contract 2017CC72MK003.
This work is also partially supported by the KU Leuven C1 grant ZKD1118 C16/16/005. 

\appendix

\section{Field redefinitions}\label{app:redefinitions}

When introducing higher-derivative corrections, the metric is ambiguous under field redefinitions $g_{\mu\nu} \rightarrow g_{\mu\nu} + X_{\mu\nu}$, where $X_{\mu\nu}$ is an object created from (at least two) metric derivatives such as in (\ref{eq:fieldredefs}). Such redefinitions shift the metric and the (higher-derivative) Lagrangian. In Appendix \ref{app:generalFR}, we will show that \emph{any} vacuum solution of (\ref{eq:EFTofGR}) and (\ref{quadraticaction}) admitting an ACMC expansion (\ref{eq:theACMC}) has multipoles that are independent of field redefinitions. For reference, we also discuss the list of all possible field redefinitions in Appendix \ref{app:specificFR}.

\subsection{Proof of invariance of multipoles}\label{app:generalFR}
We can give a general proof to show that field redefinitions do not affect the gravitational multipoles of a stationary solution to the higher-derivative theory.

A field redefinition of this metric must be of the form $g_{\mu\nu}\rightarrow g_{\mu\nu} + X_{\mu\nu}$ for some $X_{\mu\nu}$ that involves at least two derivatives acting on metric tensors, such as in (\ref{eq:fieldredefs}). It is clear that (every term in) $X_{\mu\nu}$ must either:
\begin{itemize}
 \item[a.] Contain more than one Riemann tensor (with or without derivatives acting on these Riemann tensors); or:
 \item[b.] be proportional to either the Ricci tensor or Ricci scalar.
\end{itemize}
Examples of (a) include everything listed below in (\ref{eq:RicciHDterms6D}) and (\ref{eq:listeventerms}); 
(b) is essentially the special cases $R_{\mu\nu}$ and $g_{\mu\nu} R$.

We start with a metric brought to the ACMC-form (\ref{eq:theACMC}).
Let us rewrite this as:
\be \label{eq:metricACMCS}
 g_{00} = -1 + \sum_{\ell=0}^\infty \frac{ \mathcal{S}_{\ell}}{r^{\ell+1}},\qquad
 g_{0i} = \sum_{\ell=1}^\infty \frac{\mathcal{S}_\ell}{r^{\ell+1}},\qquad
 g_{ij} = \delta_{ij} + \sum_{\ell=0}^\infty \frac{\mathcal{S}_\ell}{r^{\ell+1}},
 \ee
 where $\mathcal{S}_\ell$ is short for ``any angular dependence up to and including the order $\ell$ spherical harmonics''. For example, in $g_{00}$, this includes the leading-order contribution from the multipoles $M_\ell$, but also the (gauge-dependent) subleading terms proportional to the coefficients $c_{\ell\ell'}^{(tt)}$.
 
Two properties of such angular dependences are important:\footnote{These are easiest to understand and derive using STF tensors; see e.g. \cite{Thorne:1980ru,Mayerson:2022ekj}. Similar arguments were used in deriving the multipole structure of almost-BPS microstate geometries in \cite{Bah:2021jno}.}
\begin{itemize}
 \item[(i)] Derivatives do not increase (maximal) angular dependence\footnote{Derivatives may decrease the maximal angular dependence depending on their index structure.}, i.e.
 \be \partial_i\partial_j\cdots \partial_k \left( \frac{ \mathcal{S}_{\ell}}{r^{\ell+1}}\right) = \sum_{\ell'} \frac{\mathcal{S}_{\ell'}}{r^{\ell'+1}},\ee
 where the particulars of the sum over $\ell'$ depend on the index structure in $ij\cdots k$.
 \item[(ii)] Multiplying two ``leading'' terms gives ``subleading'' terms, i.e.:
 \be \left(\sum_{\ell=1}^\infty \frac{ \mathcal{S}_{\ell}}{r^{\ell+1}}\right)\left(\sum_{\ell'=1}^\infty \frac{ \mathcal{S}_{\ell'}}{r^{\ell'+1}}\right) = \sum_{\ell''=1}^\infty \frac{ \mathcal{S}_{\ell''-1}}{r^{\ell''+1}}.\ee
\end{itemize}
It is quite easy to see that these properties imply that the inverse metric is also given by (\ref{eq:metricACMCS}) where all the indices are simply written raised.
 
To show that the multipoles are not affected by a field redefinition, we need to show that:
\be X_{\mu\nu} = \sum_{\ell=1}^\infty \frac{ \mathcal{S}_{\ell-1}}{r^{\ell+1}},\ee
where we note the subscript $\ell-1$ on the angular dependence $\mathcal{S}$ ensures that the $\ell$-th multipole (either $M_\ell$ or $S_\ell$) is unaffected by the shift $X_{\mu\nu}$. Since a Riemann tensor involves two derivatives of the metric, it is not hard to see (using property (i)) that every component satisfies:
\be R_{\mu\nu\rho\sigma} = \sum_{\ell=1}^\infty \frac{ \mathcal{S}_{\ell}}{r^{\ell+1}}.\ee
Then, using property (ii), we immediately have that the product of two Riemann tensors, no matter what indices are involved, will always be subleading:
\be R_{\mu\nu\rho\sigma}R_{\alpha\beta\gamma\delta} = \sum_{\ell=1}^\infty \frac{ \mathcal{S}_{\ell-1}}{r^{\ell+1}}, \ee
so that it is clear all possible field redefinitions under (a) will not affect the gravitational multipoles.

For the Ricci tensor and Ricci scalar shifts under (b), a bit of calculation shows that:
\be R_{00} = \sum_{\ell=1}^\infty \frac{ \mathcal{S}_{\ell-1}}{r^{\ell+1}}, \qquad R_{0i} = \sum_{\ell=1}^\infty \frac{ \mathcal{S}_{\ell-1}}{r^{\ell+1}},\ee
but that:
\be R_{ij} = \frac12 \partial_i\partial_j g_{00} + \frac12\left( \partial_i\partial_k g_{jk} + \partial_j\partial_k g_{ik} - \partial_i\partial_j g_{kk}\right) +     \sum_{\ell=1}^\infty \frac{ \mathcal{S}_{\ell-1}}{r^{\ell+1}},\ee
where the sum over the repeated index $k$ is implied. This also means that:
\be R = \partial_i \partial_k g_{ik} + \sum_{\ell=1}^\infty \frac{ \mathcal{S}_{\ell-1}}{r^{\ell+1}}.\ee
So the Ricci tensor and scalar are not necessarily subleading --- for general spacetimes, these field redefinitions under (b) could in principle shift the multipole structure, making it ill-defined under field redefinitions. However, for \emph{vacuum} solutions, the leading order, two-derivative solution has $R_{\mu\nu}=R=0$ and hence, with the help of (i), these field redefinitions trivially also do not affect the gravitational multipoles.

Therefore, for any solutions to the Lagrangian (\ref{eq:EFTofGR}), and for vacuum solutions to (\ref{quadraticaction}) (i.e. where $\phi_{1,2}\sim \mathcal{O}(\ell)$), we can conclude that there are no possible field redefinitions that can alter the multipole structure. This, of course, includes the (higher-derivative corrected) Kerr solution we consider in this paper.

\subsection{Possible field redefinitions}\label{app:specificFR}

The previous section provided a general proof to show that the multipole moments are invariant under field redefinitions. Here, we will list and discuss the possible field redefinitions in more detail.
As we mentioned in the main text, the effective action contains all of the independent terms that contain pure Riemann (or equivalently Weyl) curvature, but not Ricci curvature \req{eq:EFTofGR}. 
The rest of the terms have been implicitly removed by using field redefinitions. However, non-linear field redefinitions can still be introduced by terms linear in the Ricci curvature.

In the six-derivative Lagrangian there are only two independent terms which are linear in Ricci curvature, and these can be chosen as

\begin{equation}\label{eq:RicciHDterms6D}
R\,\mathcal{C}\, ,\quad  R\,\tilde{\mathcal{C}}\, .
\end{equation}
A linear combination of these terms in the Lagrangian,

\begin{equation}
{\cal L}_{\text{eff}}\supset \beta_1\,R\,{\cal C}+\beta_2\,R\,{\tilde{\cal C}}\,, 
\end{equation}
can be cancelled by the following (perturbative) field redefinition,

\begin{equation}\label{eq:6dervFR}
g_{\mu\nu}\rightarrow g_{\mu\nu}\left[1+\ell^4\left(\beta_1\mathcal{C}+\beta_2\tilde{\mathcal{C}}\right)\right]\, .
\end{equation}
Note that these terms involve the product of two Riemann tensors, and clearly do not affect the multipole moments due to the arguments given above. Hence, \eqref{eq:6dervFR} preserves the multipole structure.

Using the results in \cite{Fulling_1992}, we can find the list of corresponding eight-derivative invariants. For the even-parity invariants we have
\begin{equation}\label{eq:listeventerms}
\begin{aligned}
&R\tensor{R}{_{\mu\nu }^{\rho\sigma}}\tensor{R}{_{\rho\sigma }^{\delta\gamma}}\tensor{R}{_{\delta\gamma }^{\mu\nu }}\, ,\quad \nabla^{\alpha}R R^{\mu\nu\rho\sigma}\nabla_{\alpha}R_{\mu\nu\rho\sigma}\, ,\quad R \nabla^{\alpha} R^{\mu\nu\rho\sigma}\nabla_{\alpha}R_{\mu\nu\rho\sigma}\, ,\\
&R^{\alpha\beta} \nabla_{\alpha} R^{\mu\nu\rho\sigma}\nabla_{\beta}R_{\mu\nu\rho\sigma}\, ,\quad
\nabla^{\alpha}R^{\mu\nu}\tensor{R}{^{\rho\sigma}_{\mu\beta}}\nabla^{\beta}R_{\rho\sigma\nu\alpha}\, ,\quad \nabla^{\alpha}\nabla^{\beta}\nabla^{\mu}R^{\nu\sigma}\nabla_{\mu}R_{\nu\alpha\sigma\beta}\, ,\\
&\nabla^{\mu}\nabla^{\nu}\nabla^{\rho}\nabla^{\sigma} R R_{\mu\nu\rho\sigma}\, .
\end{aligned}
\end{equation}
Integrating by parts, using Bianchi and Ricci identities and ignoring terms with more than one Ricci curvature, one can show that a linear combination of these terms gives rise to the following terms in the effective Lagrangian,

\begin{equation}
{\cal L}_{\rm{eff}}\supset \ell^{6}G_{\mu \nu}\left[K^{\mu\nu}-g^{\mu\nu}\left(L+\tfrac{1}{2}K\right)\right]\,,
\end{equation}
where

\begin{equation}
K_{\mu\nu}=\gamma_{1}\,\nabla_{(\mu|}R_{\alpha\beta\gamma\delta}\nabla_{|\nu)}R^{\alpha\beta\gamma\delta}+\,\gamma_{2}R_{\alpha\rho\sigma\beta}\nabla^{\alpha}\nabla^{\beta}R_{(\mu|}{}^{\rho\sigma}{}_{|\nu)}\,,
\end{equation}
and

\begin{equation}
L=\gamma_{3}\,R_{\mu\nu\rho\sigma}R^{\rho\sigma}{}_{\alpha\beta}R^{\alpha\beta\mu\nu}+\gamma_{4}\,\nabla_{\alpha}R_{\mu\nu\rho\sigma}\nabla^{\alpha}R^{\mu\nu\rho\sigma}+\gamma_{5}\, R^{\mu\nu\rho\sigma}\nabla^2R_{\mu\nu\rho\sigma}\, .
\end{equation}
These terms are cancelled by the following field redefinition

\begin{equation}
g_{\mu\nu}\to g_{\mu\nu}+\ell^{6}\left[g_{\mu\nu}\left(L+\tfrac{1}{2}K\right)-K_{\mu\nu}\right]\, .
\end{equation}
These terms all involve the produce of at least two Riemann tensors, and so again do not affect the multipole moments. To give an explicit example, consider $K_{\mu\nu}$, evaluated on the Kerr solution in Boyer-Lindquist coordinates and expanded in $1/r$:
\begin{equation}
\begin{aligned}
K_{tt}=\,&\frac{18 \gamma_{2} M^3}{r^9}\left[1-\frac{2M}{r}-\frac{19 M^2 \chi ^2 x^2-2 M^2 \chi ^2}{r^{2}}+\dots\right]\, ,\\
K_{t\phi}=\,&-\frac{36\gamma_2 M^4 (1-x^2)\chi}{r^9}\left[1-\frac{M}{r}+\frac{M^2 \chi ^2-18 M^2 \chi^2x^2}{r^{2}}+\dots\right]\, ,
\end{aligned}
\end{equation}
These terms clearly do not contribute to any leading order, multipole coefficient --- they will only contribute to the $c_{\ell\ell'}$ coefficients in (\ref{eq:theACMC}).

Finally, for parity-odd terms, we can obtain a basis by replacing one of the Riemann tensors in \eqref{eq:listeventerms} by its dual. The rest of the analysis is analogous, and these terms also do not give any contribution to the multipoles.


\section{Surface charges}\label{app:Waldformalism}

In this appendix we compute the Iyer-Wald 2-form ${\bf k}_{\xi}$ for a general class of higher-derivative theories which include as particular cases those considered in this work. 

\subsection{Some generalities}

Let us consider a class of theories characterized by the following $d$-form Lagrangian

\begin{equation}
 {\mathbf L}={\boldsymbol\epsilon}\, {\mathcal L}(R_{\mu\nu\rho\sigma}, g^{\alpha\beta}, \partial_{\mu}\Phi, \Phi)\,,   
\end{equation}
where $\Phi=\{\phi^{A}\}_{A=1, \dots, N}$ is a set of (pseudo)scalar fields and

\begin{equation}
{\boldsymbol\epsilon}=\frac{1}{d!}\epsilon_{\mu_1 \dots \mu_d}\, dx^{\mu_1}\wedge \dots \wedge dx^{\mu_d}=\sqrt{-g}\, d^dx\, ,
\end{equation}
is the volume form. In what follows, we will make use of the following notation 

\begin{equation}
{\boldsymbol{\epsilon}}_{\mu_1\dots \mu_n}= \frac{1}{(d-n)!}\epsilon_{\mu_1\dots \mu_n \nu_1 \dots \nu_{d-n}}\, dx^{\nu_1 }\wedge \dots \wedge dx^{\nu_{d-n}} \, .
\end{equation}
Under general variations of the fields, we have that 

\begin{equation}\label{eq:dL}
\delta {\bf L}={\boldsymbol\epsilon} \left({\cal E}_{\mu\nu}{\delta g}^{\mu\nu}+{\cal E}_{A} \delta \phi^{A}\right)+d {\boldsymbol\Theta} (\varphi, \delta\varphi)\,,
\end{equation}
where $\varphi=\left\{g_{\mu\nu}, \Phi\right\}$ denotes schematically all the dynamical fields of the theory (metric and scalars), ${\boldsymbol\Theta}$ is the boundary term that arises upon integration by parts and 

\begin{equation}
\sqrt{-g}\,{\cal E}_{\mu\nu}\equiv \frac{\delta S}{\delta g^{\mu\nu}}\, ,\hspace{1cm}\sqrt{-g}\,{\cal E}_{A}\equiv \frac{\delta S}{\delta \phi^{A}}\, .
\end{equation}
Hence, the field equations are ${\cal E}_{\mu\nu}=0$, ${\cal E}_{A}=0$.\footnote{We assume ${\cal E}_{\mu\nu}$ is symmetric by construction.} For the class of theories under consideration, we find 

\begin{equation}\label{eq:variationL}
\begin{aligned}
\delta {\bf L}=\,&{\boldsymbol\epsilon}\left\{{\delta g}^{\mu\nu}\left(\frac{\partial {\cal L}}{\partial g^{\mu\nu}}-\frac{1}{2}g_{\mu\nu}\,{\cal L}-P^{\rho\sigma\lambda}{}_{\mu}R_{\rho\sigma\lambda\nu}-2\nabla^{\alpha}\nabla^{\beta}P_{\beta\mu\nu\alpha}\right)\right.\\[1mm]
&\left.+\delta \phi^{A}\left(-\nabla_{\mu}P_{A}^{\mu}+\frac{\partial {\cal L}}{\partial \phi^{A}}\right)+\nabla_{\mu}\theta^{\mu}\right\}\,,
\end{aligned}
\end{equation}
where 

\begin{equation}
\theta^{\mu}=P^{\mu}_{A}\delta \phi^{A}-2P^{\mu\nu\rho}{}_{\sigma}\delta \Gamma^{\sigma}_{\nu\rho}-2\nabla_\sigma P^{\mu\nu\rho\sigma}\delta g_{\nu\rho}\,,
\end{equation}
and 

\begin{equation}
P^{\mu\nu\rho\sigma}\equiv\frac{\partial {\cal L}}{\partial R_{\mu\nu\rho\sigma}}\,, \hspace{1cm} P^{\mu}_{A}\equiv\frac{\partial {\cal L}}{\partial (\partial_{\mu}\phi^{A})}\,.
\end{equation}
Eq.~\eqref{eq:variationL} can be further massaged using an identity that relates $P^{\mu\nu\rho\sigma}$, $P^{\mu}_{A}$ and $\frac{\partial {\cal L}}{\partial g^{\mu\nu}}$ and which can be derived using the fact that the Lie derivative of the Lagrangian $\cal L$ can be written in two different ways  \cite{Padmanabhan:2011ex}. First, as 

\begin{equation}\label{eq:LieL1}
{\mathsterling}_{\xi}{\cal L}=\xi^{\alpha}\partial_{\alpha}{\cal L}=\xi^{\alpha}\left(P^{\mu\nu\rho\sigma}\nabla_{\alpha}R_{\mu\nu\rho\sigma}+P^{\mu}_{A}\nabla_{\alpha}\partial_{\mu}\phi^{A}+\frac{\partial {\cal L}}{\partial \phi^A}\partial_{\alpha}\phi^A\right)\, .
\end{equation}
And second, using the chain rule:

\begin{equation}\label{eq:LieL2}
{\mathsterling}_{\xi}{\cal L}=P^{\mu\nu\rho\sigma}{\mathsterling}_{\xi}R_{\mu\nu\rho\sigma}+\frac{\partial {\cal L}}{\partial g^{\mu\nu}}{\mathsterling}_{\xi}g^{\mu\nu}+P^{\mu}_{A}{\mathsterling}_{\xi}\partial_{\mu}\phi^A+\frac{\partial {\cal L}}{\partial \phi^A}{\mathsterling}_{\xi}\phi^A\, .
\end{equation}
Expanding the Lie derivatives, we get

\begin{equation}\label{eq:Liederivativesexp}
\begin{aligned}
P^{\mu\nu\rho\sigma}{\mathsterling}_{\xi}R_{\mu\nu\rho\sigma}=\,&P^{\mu\nu\rho\sigma}\xi^{\alpha}\nabla_{\alpha}R_{\mu\nu\rho\sigma}+4 P^{\rho\sigma\lambda}{}_{\mu}R_{\rho\sigma\lambda \nu}\nabla^{\mu}\xi^{\nu}\,,\\[1mm]
\frac{\partial {\cal L}}{\partial g^{\mu\nu}}{\mathsterling}_{\xi}g^{\mu\nu}=\,&-2\frac{\partial {\cal L}}{\partial g^{\mu\nu}}\nabla^{(\mu}\xi^{\nu)}\,,\\[1mm]
P^{\mu}_{A}{\mathsterling}_{\xi}\partial_{\mu}\phi^{A}=\,&P^{\mu}_{A}\xi^{\alpha}\nabla_{\alpha}\partial_{\mu}\phi^{A}+ P_{A\,\mu}\partial_{\nu}\phi^{A} \,\nabla^{\mu}\xi^{\nu}\,,\\[1mm]
\frac{\partial {\cal L}}{\partial \phi^A}{\mathsterling}_{\xi}\phi^A=\,&\frac{\partial {\cal L}}{\partial \phi^A}\xi^{\alpha}\partial_{\alpha}\phi^A\,.
\end{aligned}
\end{equation}
Plugging \eqref{eq:Liederivativesexp} into \eqref{eq:LieL2} and using \eqref{eq:LieL1}, we arrive at the following identity

\begin{equation}
\nabla^{(\mu}\xi^{\nu)}\left(4 P^{\rho\sigma\lambda}{}_{\mu}R_{\rho\sigma\lambda \nu}-2\frac{\partial {\cal L}}{\partial g^{\mu\nu}}+ P_{A\,\mu}\partial_{\nu}\phi^A\right)+\nabla^{[\mu}\xi^{\nu]}\left(4 P^{\rho\sigma\lambda}{}_{\mu}R_{\rho\sigma\lambda \nu}+ P_{A\,\mu}\partial_{\nu}\phi^A\right)=0\,,
\end{equation}
from which we deduce that 

\begin{eqnarray}
\label{eq:id1}
\frac{\partial {\cal L}}{\partial g^{\mu\nu}}&=\,&2 P^{\rho\sigma\lambda}{}_{(\mu|}R_{\rho\sigma\lambda |\nu)}+ \frac{1}{2}P_{A\,(\mu}\partial_{\nu)}\phi^{A}\,,\\[1mm]
\label{eq:id2}
P_{A\,[\mu}\partial_{\nu]}\phi^A&=\,&-4 P^{\rho\sigma\lambda}{}_{[\mu|}R_{\rho\sigma\lambda |\nu]}\, .
\end{eqnarray}
In absence of scalars, the left-hand side of \eqref{eq:id2} vanishes and we have that the tensor $P^{\rho\sigma\lambda}{}_{\mu}R_{\rho\sigma\lambda \nu}$ is totally symmetric under the exchange of the free indices. This will also be the case for the theories we are interested in, since $P_{A\,\mu}\partial_{\nu}\phi^{A}=-\partial_{\mu}\phi^{A}\partial_{\nu}\phi^{B}\delta_{AB}$. Hence, we shall assume this property in what follows. Let us then use \eqref{eq:id1} in \eqref{eq:variationL} to finally write the variation of $\bf L$ as 

\begin{equation}\label{eq:deltaL}
\begin{aligned}
\delta {\bf L}=\,&{\boldsymbol\epsilon}\left\{{\delta g}^{\mu\nu}\left(-\frac{1}{2}g_{\mu\nu}\,{\cal L}+P^{\rho\sigma\lambda}{}_{\mu}R_{\rho\sigma\lambda\nu}-2\nabla^{\alpha}\nabla^{\beta}P_{\beta(\mu\nu)\alpha}+\frac{1}{2}P_{A\,\mu}\partial_{\nu}\phi^A\right)\right.\\[1mm]
&\left.+\delta \phi^A\left(-\nabla_{\mu}P^{\mu}_{A}+\frac{\partial {\cal L}}{\partial \phi^A}\right)+\nabla_{\mu}\theta^{\mu}\right\}\,.
\end{aligned}
\end{equation}
Comparing with \eqref{eq:dL}, we can read off ${\cal E}_{\mu\nu}$ and ${\cal E}_{A}$, 

\begin{eqnarray}
{\cal E}_{\mu\nu}&=\,&-\frac{1}{2}g_{\mu\nu}\,{\cal L}+P^{\rho\sigma\lambda}{}_{\mu}R_{\rho\sigma\lambda\nu}-2\nabla^{\alpha}\nabla^{\beta}P_{\beta(\mu\nu)\alpha}+\frac{1}{2}P_{A\,\mu}\partial_{\nu}\phi^{A}\,, \\[1mm]
{\cal E}_{A}&=\,&-\nabla_{\mu}P^{\mu}_{A}+\frac{\partial {\cal L}}{\partial \phi^{A}}\, ,
\end{eqnarray}
as well as the boundary term ${\boldsymbol\Theta}$, which is given by 

\begin{equation}
{\boldsymbol\Theta}={\boldsymbol\epsilon}_{\mu}\theta^{\mu}={\boldsymbol\epsilon}_{\mu}\left(P^{\mu}_{A}\delta \phi^{A}-2P^{\mu\nu\rho}{}_{\sigma}\delta \Gamma^{\sigma}_{\nu\rho}-2\nabla_\sigma P^{\mu\nu\rho\sigma}\delta g_{\nu\rho}\right)\,.
\end{equation}

\subsection{Noether charge}

Let us consider the variation of ${\bf L}$ under diffeomorphisms generated by a vector field $\xi^{\mu}$. The variations of the fields are given by

\begin{equation}
\begin{aligned}
\delta_{\xi}g_{\mu\nu}=&\mathsterling_{\xi} g_{\mu\nu}=2\nabla_{(\mu}\xi_{\nu)}\,, \\[1mm]\delta_{\xi}\Phi=&\mathsterling_{\xi} \Phi=\xi^{\mu}\partial_{\mu}\Phi\, .
\end{aligned}
\end{equation}
Therefore, 

\begin{equation}\label{eq:deltaxiL}
\begin{aligned}
\delta_{\xi}{\bf L}=\,&\mathsterling_{\xi} {\bf L}=d\left(\iota_{\xi}{\bf L}\right)\\[1mm]
=\,&{\boldsymbol\epsilon} \left(-2\,{\cal E}_{\mu\nu}\nabla^{\mu}\xi^{\nu}+{\cal E}_{A}\xi^{\nu} \partial_{\nu}\phi^{A}\right)+d {\boldsymbol\Theta} (\varphi, \delta\varphi)\\[1mm]
=\,&{\boldsymbol\epsilon} \left[-2\nabla^{\mu}\left(\,{\cal E}_{\mu\nu}\xi^{\nu}\right)+2\nabla^{\mu}{\cal E}_{\mu\nu}\xi^{\nu}+{\cal E}_{A} \xi^{\nu} \partial_{\nu}\phi^{A}\right]+d {\boldsymbol\Theta} (\varphi, \delta\varphi)\\[1mm]
=\,&d\left[{\boldsymbol\Theta} (\varphi, \delta\varphi)-\xi^{\mu}{\bf C}_{\mu}\right]\,,
\end{aligned}
\end{equation}
where in the last line we have made use of the Noether identity 

\begin{equation}
2\nabla^{\mu}{\cal E}_{\mu\nu}=-{\cal E}_{A} \partial_{\nu}\phi^{A}\, ,
\end{equation}
and we have defined 

\begin{equation}
{\bf C}_{\mu}=2\epsilon_{\nu}{\cal E}^{\nu}{}_{\mu}\, .
\end{equation}
Following \cite{Wald:1993nt, Iyer:1994ys}, we can associate a Noether current $(d-1)$-form to the vector $\xi$ as follows

\begin{equation}
{\bf j}_{\xi}\equiv{\boldsymbol\Theta} (\varphi, \delta\varphi)-\iota_{\xi}{\bf L}\,.
\end{equation}
From \eqref{eq:deltaxiL}, it follows that

\begin{equation}\label{eq:Noethercurrent}
{\bf j}_{\xi}= d{\bf Q}_{\xi}+\xi^{\mu}{\bf C}_{\mu}\, ,
\end{equation}
which holds off-shell and for arbitrary vector fields. $\bf{Q}_{\xi}$ is the Noether charge $(d-2)$-form, which gives the Noether charge once it is integrated over a (closed) codimension-2 hypersurface. The expression of ${\bf Q}_{\xi}$ for the theories of interest can be readily found using previous results in the literature \cite{Bueno:2016ypa, Ortin:2021ade}, since it turns out that the combination 

\begin{equation}
\begin{aligned}
{\bf j}_{\xi}-\xi^{\mu}{\bf C}_{\mu}=\,&\epsilon_{\mu}\left[4P^{\mu\nu\sigma\rho}\nabla_{\rho}\nabla_{(\nu}\xi_{\sigma)}-4 \nabla_{\sigma}P^{\mu\nu\rho\sigma}\nabla_{(\nu}\xi_{\rho)}-2P^{\rho\sigma\lambda \mu}R_{\rho\sigma\lambda}{}^{\nu}\xi_{\nu}\right.\\[1mm]
&\left.+4\nabla_{\alpha}\nabla_{\beta}P^{\beta(\mu\nu)\alpha}\xi_{\nu}\right]\,,
\end{aligned}
\end{equation}
is exactly the same as in ${\cal L}\left(R_{\mu\nu\rho\sigma}, g^{\alpha\beta}\right)$ theories.  Then, the expression for the Noether charge $(d-2)$-form is

\begin{equation}
{\bf Q}_{\xi}={\epsilon}_{\mu\nu}\left(-P^{\mu\nu\rho\sigma}\nabla_{\rho}\xi_{\sigma}+2\nabla_{\rho}P^{\mu\nu\rho\sigma}\xi_{\sigma}\right)\,,
\end{equation}
where we have assumed (without loss of generality) that 

\begin{equation}
P^{\mu[\nu\rho\sigma]}=0\, .
\end{equation}

\subsection{Surface charge}

Finally, we can define the Iyer-Wald $(d-2)$-form ${\bf k}_{\xi}$ as 

\begin{equation}
{\bf k}_{\xi}\equiv \delta {\bf Q}_{\xi}- \iota_{\xi}{\boldsymbol\Theta} \left(\varphi,\delta\varphi\right)\, .
\end{equation}
Using previous definitions, one can check that on-shell and whenever $\delta \varphi$ satisfies the linearized equations of motion, we have that

\begin{equation}
{\boldsymbol\omega}(\varphi,\delta \varphi, {\mathsterling}_{\xi}\varphi)=d{\bf k}_{\xi}\,,
\end{equation}
where ${\boldsymbol\omega}(\varphi, \delta_1\varphi, \delta_2\varphi)$ is the pre-symplectic current $(d-1)$-form, defined as follows

\begin{equation}
{\boldsymbol\omega}(\varphi, \delta_1\varphi, \delta_2\varphi)\equiv \delta_1 {\boldsymbol\Theta}\left(\varphi, \delta_2\varphi\right)-\delta_2 {\boldsymbol\Theta}\left(\varphi, \delta_1\varphi\right)\, .
\end{equation}
The expression of ${\bf k}_{\xi}$ for the theories under consideration is

\begin{equation}
\begin{aligned}
{\bf k}_{\xi}=\,&{\boldsymbol \epsilon}_{\mu\nu}\left[-\delta P^{\mu\nu\rho}{}_{\sigma}\nabla_{\rho}\xi^{\sigma}-P^{\mu\nu\rho\sigma}\nabla_{\rho}\delta g_{\sigma \lambda}\xi^{\lambda}+2 \delta \left(\nabla_{\rho}P^{\mu\nu\rho}{}_{\sigma}\right)\xi^{\sigma}\right.\\[1mm]
&\left.\frac{1}{2}\left(-P^{\mu\nu\rho\sigma}\nabla_{\rho}\xi_{\sigma}+2\nabla_{\rho}P^{\mu\nu\rho\sigma}\xi_{\sigma}\right)g^{\alpha\beta}\delta g_{\alpha\beta}\right.\\[1mm]
&\left.-\xi^{\nu}\left(P^{\mu}_{A}\delta \phi^{A}+2P^{\mu\alpha\beta \rho}\nabla_{\rho}\delta g_{\alpha\beta}-2\nabla_\sigma P^{\mu\alpha\beta\sigma}\delta g_{\alpha\beta}\right)\right]\, .
\end{aligned}
\end{equation}

\bibliographystyle{JHEP}
\bibliography{Gravities.bib}

\end{document}